\documentclass[5p]{elsarticle}

\usepackage{lineno}
\modulolinenumbers[5]
\usepackage{hyperref}
\usepackage{color}
\usepackage[figuresleft]{rotating}
\usepackage{caption}
\usepackage{subcaption}
\usepackage{layout}
\usepackage{setspace}
\usepackage{tabularx}

\journal{ }

\usepackage[utf8]{inputenc}
\usepackage{mathtools}
\usepackage{color}
\usepackage{listings}

\lstdefinelanguage{GAMS}{
morekeywords={
ABORT , ACRONYM , ACRONYMS , ALIAS , model , AND , ASSIGN , BINARY , CARD , DISPLAY , EPS , EQ , EQUATION , EQUATIONS , GE , GT , INF , INTEGER , LE , LOOP , LT , MAXIMIZING , MINIMIZING , MODEL , NA , NE , NEGATIVE , NOT , OPTION , OPTIONS , OR , PARAMETERS , POSITIVE , PROD , SCALAR , SCALARS , SET , SETS , SMAX , SMIN , SOS1 , SOS2 , SYSTEM , TABLE , USING , VARIABLE , VARIABLES , XOR , YES , REPEAT , UNTIL , WHILE , IF , THEN , ELSE , SEMICONT , SEMIINT , FILE , FILES , PUT , PUTPAGE , PUTTL , PUTCLOSE , FREE , NO , SOLVE , ELSEIF , ABS , ARCTAN , CEIL , COS , ERROR , EXP , FLOOR , LOG , LOG10 , MAP , MAPVAL , MAX , MIN , MOD , NORMAL , POWER , ROUND , SIGN , SIN , SQR , SQRT , TRUNC , UNIFORM , LO , UP , FX , PRIOR , PC , PS , PW , TM , BM , CASE , DATE , IFILE , OFILE , PAGE , RDATE , RFILE , RTIME , SFILE , TITLE , TS , TL , TE , TF , LJ , NJ , SJ , TJ , LW , NW , SW , TW , ND , NR , NZ , CC , HDCC , TLCC , LL , HDLL , TLLL , LP , WS , /,PROD: },
sensitive = false,
morecomment=[f]*, 
morecomment=[s]{$ontext}{$offtext},
morecomment=[s][\color{black}]{/}{/},
}
\usepackage{setspace}
\usepackage{tabularx} 
\newcolumntype{Y}{>{\centering\arraybackslash}X}
\usepackage[misc,geometry]{ifsym}
\usepackage{natbib}

\biboptions{authoryear}

\makeatletter
\def\ps@pprintTitle{%
 \let\@oddhead\@empty
 \let\@evenhead\@empty
 \def\@oddfoot{}%
 \let\@evenfoot\@oddfoot}
\makeatother

\begin{document}

\begin{frontmatter}

\title{Optimal Population in a Finite Horizon}
\author[]{Satoshi Nakano}
\author[]{Kazuhiko Nishimura\corref{cor1}}
\ead{nishimura@n-fukushi.ac.jp\,}


\begin{abstract}
A favorable population schedule for the entire potential human family is sought, under the overlapping generations framework, by treating population (or fertility) as a planning variable in a dynamical social welfare maximization context. 
The utilitarian and maximin social welfare functions are examined, with zero future discounting, while infinity in the maximand is circumvented by introducing the depletion of energy resources and its postponement through technological innovations. 
The model is formulated as a free-horizon dynamical planning problem, solved via a non-linear optimizer.
Under {{exploratory}} scenarios, we visualize the potential trade-offs between the two welfare criteria.
\end{abstract}

\begin{keyword}
Population \sep Overlapping Generations Model \sep Resources Depletion \sep Intergenerational Equity \sep Numerical Methods \end{keyword}

\end{frontmatter}


\section{Introduction}

One of the fundamental sources of economic growth 
can be found in the dynamics of human population.
The models embodying various growth theories virtually employ human population and technological progression as the fundamental driving forces of growth.
One may further conclude that technological progress being endogenous, as assumed in the new growth theory literature, ultimately attributes to the escalation of labor productivity.
It is then not surprising that the problem of economic growth verses ecological sustainability has been discussed in relation to the explosive increase in human population \citep[e.g., ][]{ehrlich71, Holdren08Science}.
In most of the previous dynamic macroeconomic models, however, it has been assumed that population and its growth is determined by the course of nature.

With the classic Ramseyan models in general, future generations are represented by a single infinitely lived agent (ILA) so that one may set aside the scale effect of population (or its growth) when examining welfare.
For the overlapping generations (OLG) models, which capture the behaviors of different generations overlapped within the finite lengths of their lifetimes, the fertility rate can be either exogenously given or conducted endogenously.\footnote{
\citet{bb} considered incentives (altruism) and the costs of child-raising, thus endogenizing fertility in the study of economic growth.
{{{\citet{ecks} assumed individuals include parenthood in their utilities under the OLG framework.
}}}
}
ILA models have been used in the major climate change debates including \citet{stern} and \citet{nord}, while
the OLG models \citep[e.g.,][]{HN1992AER, ecoeco2001} have also been viewed as standard and their performances have been compared with those of the ILA models \citep{stephan, howarth2000ERE, EER2012}.

In contrast to the preveous
models, our interest is in investigating the optimal population schedule by taking the fertility rate as a planning variable (as we conform to \citet{seps} and \citet{reinke}) with respect to a social welfare function (SWF) from a social planner's point of view.
Moreover, we are interested in the intergenerational equity upon 
planning, virtually treating all (potential) people in an equitable manner.
It is known, however, that Pareto efficiency and anonymity (intergenerational equity) in aggregating infinite utility streams are incompatible with the ILA models.\footnote{It is known that there does not exist any such SWF satisfying the axioms of both Pareto and anonymity in the infinite streams of utility \citep{Diamond65Econometrica, BM03Econometrica}. See \hyperlink{ap1}{Appendix 1} for an exemplification.} 
A planning feature can be incorporated in the OLG models, which are prone to inefficient over-saving (dynamic inefficiency) in the infinite horizon; however, in this case as well, zero discounting and the infinite horizon are incompatible; 
\citet{gigliotti} used an OLG model with a discounted sum of the total utility per generation to investigate the optimal population policy.

Hence, in theory, our strategy for maintaining intergenerational equity in the optimal planning of population must be to relax the infiniteness of the planning horizon.
For intergenerational equity we adopt such welfarist principles as classical utilitarian (or, Benthamite) and maximin (or Rawlsian).
More specifically, 
utilitarian SWF is $\sum_i u_i$, where $u_i$ denotes the utility of an individual $i$, while maximin SWF is $\min u_i$.\footnote{In this study, we will not go further into combining these SWFs as discussed in \citet{hw} and \citet{nvl}.}
A population study under classical utilitarianism leads to a repugnant conclusion \citep{parfit}, such that utilitarian SWF is maximized at the largest possible population with the lowest level of utility.
The implication of this proposition in the dynamic context is that population should grow at the maximum rate physically possible.\footnote{\citet{renstrom} showed that critical level utilitarianism \citep{BBD} can yield an interior solution (hence, avoid repugnant conclusion) in a future-discounted infinite horizon model. \citet{shiell} showed that the boundedness of the resources can avoid a repugnant conclusion.}
Finiteness of the primary factor of production (i.e., energy), on the other hand, can avoid such a result over intergenerational equity, as we will discuss later on.

Perhaps the distinguishing feature of this study can be attributed to the endogeneity of the planning horizon.
We confront, at least in this study, the fact that any human activity can only consume energy and cannot manufacture energy without consuming more.\footnote{This is a corollary to the laws of thermodynamics. 
We will use the term \textit{energy} hereafter, in stead of available or free energy \citep{myth, ayres}, although they may be more appropriate in this regard.
Also in the same vein, we preclude \textit{dilute} energy sources such as solar, terrestrial, and lunar power, although they are persistent.
}
In that regard, human technological efforts can only postpone the final depletion of an \text{energy} stock (i.e., sources of energy such as coal, oil, natural gas, uranium, etc.) along with the stream of forthcoming generations.
Under such perspective, the planning horizon is dependent upon the trajectory of the population dynamics.
More specifically, the main target of this study will be to solve a free-horizon dynamic planning problem, whose control is the trajectory of the population (or fertility rate), maximizing a welfarist SWF under a finite amount of energy stock.\footnote{Some may argue that the availability of energy may increase indefinitely due to persistent innovation (for example, fast breeder reactor, nuclear fusion, etc.), but the possibility that we run out of energy to even exploit those technologies and that further availability being beyond the reach of the human hand, is not zero.  
This study is concerned with the latter conservative scenario.}

\begin{figure}[ht!]
\begin{centering}
\includegraphics[scale=0.6]{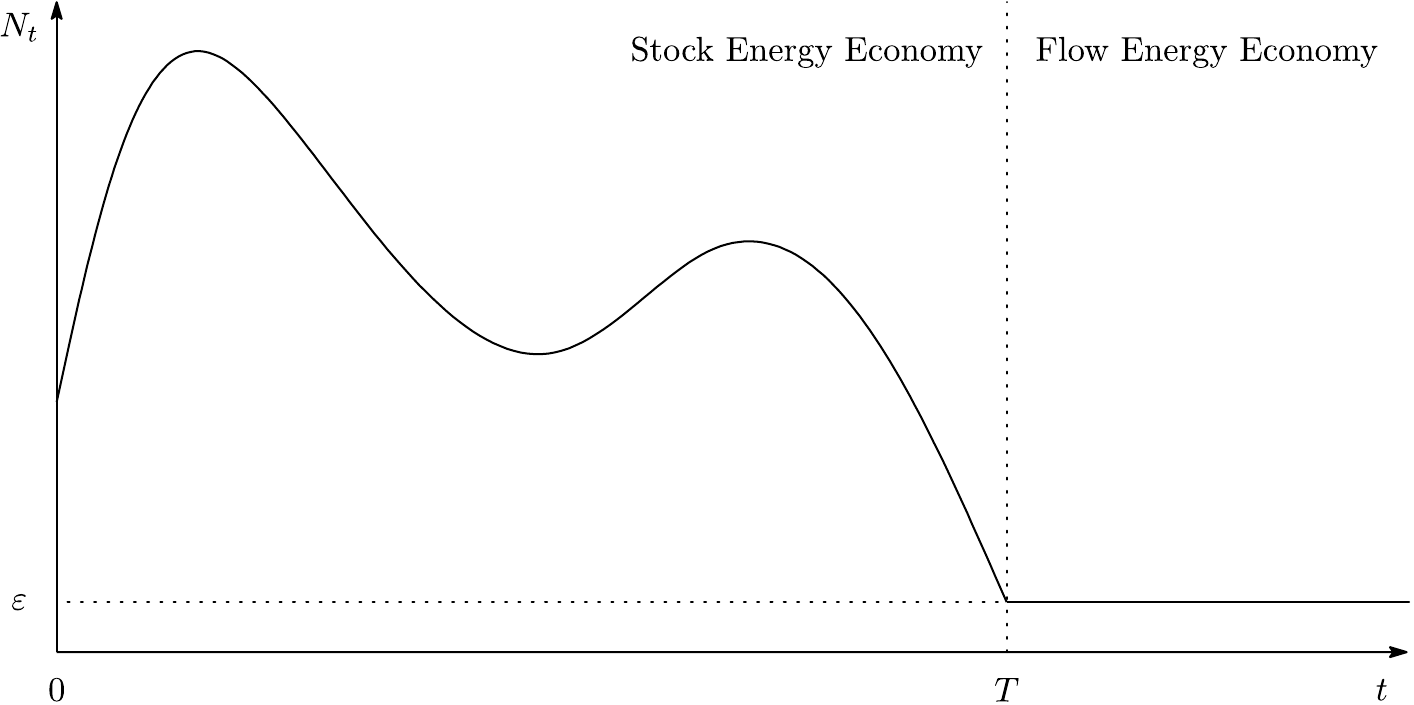}
\caption{Sample population schedule for the stock-energy economy.}\label{fig_intro}
\end{centering}
\end{figure}

Upon depletion of the energy stock, what remains will be a persistent yet dilute energy, such as hydro and wind power.
Since energy is complementary to human activity, the magnitude of the post stock energy economy will be capped by the dilute energy that is assumed to persist in perpetuity. 
This flow energy economy will have no planning involved, as the population (labor) cannot affect production over a certain threshold.
We assume a fixed level of the appropriate population of the post stock energy economy (i.e., the flow energy economy), which we denote by $\varepsilon$. 
We further assume that the final remaining capital of the stock energy economy will have no effect on $\varepsilon$.  
This allows us to focus on the planning of the population for the free-finite-horizon stock energy economy, and to isolate the problem from eternity (Figure \ref{fig_intro}).

Thus, the main functional to be investigated is the population schedule for the stock energy economy, $\left\{ N_t \right\}_{t=1}^T$, where $t$ indicates generation, or time period when generation $t$ is young.
After the final generation $T$ consumed the final remaining stock energy, it is expected that $N_t = \varepsilon$, for all $t \geq T+1$.
\footnote{A post stock energy economy is a flow energy economy whose population is $\varepsilon$ for all $t > T$.}
Hence, we may otherwise call $T$ the planning horizon.  
Because the planning involves intergenerational population planning, we use the OLG model rather than the ILA as the main driver of the (stock energy) economy.
As per normal, the OLG model's main dynamics for per-capita capital stock is affected by the population and (labor) productivity growth.
Furthermore, productivity is governed by energy efficiency and resource availability, both of which is a function of the cumulative population.\footnote{Our baseline perception of energy is that it is complementary to the factors of production. 
In contrast, \citet{mitra} and \citet{asheim} treat depletable resources as a substitutable factor for production, thus allowing an infinitely large population to substitute for the depletion of resources. }  
Hence, while energy efficiency improves, resource availability will decay until the productivity reaches zero, as the population record is cumulated.

In the following section, after briefly describing the basic OLG framework, we introduce our sub-models concerning the dynamics of energy efficiency and resource availability, in regard to the dynamics of population.
We then introduce the two SWFs along with the final generation defined in the terminal condition of the planning model.
In Section 3, we discuss the solvability of the problem formulation and provide some artifices for actually solving a problem under a decent computing environment.
We further report on the parameters employed and a GAMS code used along with the results obtained.
Concluding remarks are made in Section 4.

\section{The Model}
\subsection{Overlapping Generations Model}
We begin by specifying the basic overlapping generations model in a discrete time setting for later modification.
The essential difference between the OLG and the ILA models is that there is turnover in generations.
To simplify the analysis, the following model assumes that each individual lives for only two periods, so that there will be just two different generations, who we will refer to as the young and old, at the same period of time $t$.
Let $c_t$ and $d_{t}$ denote the consumptions of the young and old, in period $t$.
We shall also call generation $t$, the group of identical representative individuals who were born in period $t$.

The problem of a representative individual of generation $t$ is to maximize his/her lifetime utility, denoted $u_t$, depending on their lifetime consumption, $c_t$ and $d_{t+1}$, that is to:
\begin{align}
\max_{c_t, d_{t+1}} ~ u_t &= u(c_t) + \beta u(d_{t+1})     \nonumber \\
\text{s.t. } ~ c_t &= w_t - s_t, ~~ d_{t+1} = \left(1+r_{t+1} \right) s_{t}  \label{constraints}
\end{align}
Here, each generation when young earns wage $w_t$ and decides whether to consume then or to save for later life.  
The saving is denoted by $s_t$ and will have some return at a rate $r_{t+1}$ in the future\footnote{The rate of return is observed in $t$.} so that generation $t$ will receive $(1+r_{t+1}) s_t$ in the future that could be consumed when they get old.  
The time preference of the individuals is denoted by $\beta$.

The first order condition of the individual's problem is of the following:
\begin{align}
 u'(c_t) - \beta \left(1+r_{t+1} \right) u'(d_{t+1}) = 0  \label{euler}
\end{align}
Assume constant relative risk aversion (CRRA) for utility as follows:
\begin{align*}
u(c) = \frac{c^{1-\gamma}}{1-\gamma}
\end{align*}
where, $\gamma >0$.
For the case of CRRA, the Euler equation (\ref{euler}) will be reduced, with respect to the equations (\ref{constraints}) as follows:
\begin{align}
s_t = \frac{w_t}{1+\beta^{\frac{-1}{\gamma}} \left(1+r_{t+1} \right)^{1-\frac{1}{\gamma}}} \label{st}
\end{align}

The population of generation $t$ is denoted by $N_{t}$ and is subject to the following dynamics:
\begin{align}
N_{t+1} = (1+n_t) N_{t} \label{population}
\end{align}
where, $n_t$ denotes the growth rate of population of the generation $t$ which is usually fixed in advance.
Note that $N_{t}$ is the generation $t$'s population when they are young. 
By assuming inelastic labor supply, the amount of labor at period $t$ will hence be identical to the population of that generation, $N_t$.

The producers manufacture the num\'eraire good at each period $t$ under the production function $Y_t = F(K_t,  A_t N_t)$, where production is carried out by two primary inputs, namely, the capital stock $K_t$ and the labor $N_t$.
Note that in this case the labor input for the production function is being augmented by $A_t$, the effectiveness of labor.
Under this labor augmenting (Harrod neutral) production function, the behavior of production may be written as to
\begin{align}
\max_{K_{t}, N_{t}} ~ F(K_{t}, A_t N_t) - x_t K_{t} - w_t N_t     \label{firm} 
\end{align}
while, $A_t N_t$ is referred to as effective labor;  $x_t$ is the cost of unit capital, which will be specified later.
The production function $Y_{t}=F(K_t, A_t N_t)$ is assumed to have constant returns to scale with respect to $K_t$ and $N_t$, for which the profit would equal zero.
Hence what follows will always hold:
\begin{align}
Y_t = x_t K_t + w_t N_t  \label{zeroprofit}
\end{align}

Also, in such a case we may reduce the number of variables using capital stock per \textit{effective} labor, i.e., $k \equiv {K}/{AN}$, with the intensive form of production function, $y = F(K, AN)/AN = f(k)$. 
Then, (\ref{firm}) can be rewritten as to
\begin{align}
\max_{k_t} ~  f(k_t) - x_t k_t - w_t /A_t   \label{firm2}
\end{align}
whose first order condition is of the following:
\begin{align}
f(k_t) = f'(k_t) k_t + w_t/A_t, ~~~ f'(k_t) =x_t \label{ff}
\end{align}
By assuming that the type of production function is Cobb-Douglas, (\ref{ff}) becomes
\begin{align}
k_t^\alpha = \alpha k_t^{\alpha} + w_t/A_t, ~~~ \alpha k_t^{\alpha - 1} = x_t  \label{wt}
\end{align}
with $\alpha \in (0, 1)$ being the output elasticity of capital.

Looking into the market equilibrium at each period $t$ the economy will have the following macroeconomic balance between the aggregate yield $Y_t$, the aggregate consumption $C_t$, and the aggregate investment $I_t$:
\begin{align}
Y_t = C_t + I_t  \label{yield}
\end{align}
Meanwhile, capital stock will monetarily grow at the interest rate, while physically depreciate at a constant rate $\delta \in [0,1]$ but be reinforced by the investments, that is,
\begin{align}
K_{t+1} = (1+r_{t+1}) K_{t} = (1 - \delta) K_t + I_t 
\label{investment}
\end{align}
Furthermore, the aggregate investment $I_t$ by the young of generation $t$ is meant to purchase the entire existing capital $K_t$ at that moment from the old, and the return of this investment must coincide with the compensation for the capital stock rights in the next period, when the young gets old; in other words,
$(1+r_{t+1}) I_t = x_{t+1} K_{t+1}$.
Hence, by (\ref{investment}), we will know that the following relation will always hold:
\begin{align}
x_{t+1} = r_{t+1} + \delta \label{rt}
\end{align}

Aggregate consumption in each period $C_t$ is the sum of the consumptions of the two overlapped generations living at the same time, i.e., 
\begin{align}
C_t = N_t c_t + N_{t-1} d_t  \label{aggconsumption}
\end{align}
At period $t$ the old generation receives $I_t$ only to spend everything for the consumption in that period, while the young generation consumes $(w_t-s_t) N_t $ in aggregate, hence, by Equations (\ref{zeroprofit}, \ref{yield} and \ref{investment}), we shall modify (\ref{aggconsumption}) to obtain:
\begin{align}
C_t &= (w_t - s_t) N_t  + I_t \nonumber\\
Y_t - x_t K_t &= w_t N_t  - s_t N_t + K_{t+1} - (1-\delta) K_t \nonumber \\
s_t N_t &= K_{t+1} - (1-\delta)K_t  \nonumber\\
s_t N_t &= \left( \frac{\delta + r_{t+1}}{1+r_{t+1}} \right) K_{t+1} \nonumber\\
\frac{1}{\left(1+n_t \right) A_{t+1}} s_t &= \left( \frac{\delta + r_{t+1}}{1+r_{t+1}} \right) k_{t+1}  \label{dynamics}
\end{align}
Notice that the last modification is due to dividing both sides by $A_{t+1}N_{t+1}$.
Under the CRRA utility and constant returns to scale production, (\ref{dynamics}) will become as follows:
\begin{align}
s_t/{A_t} =&   \frac{w_t/{A_t}}{1+\beta^{\frac{-1}{\gamma}}\left(1+r_{t+1} \right)^{\frac{\gamma - 1}{\gamma}}} \notag 
= \frac{f(k_t) - f'(k_t) k_t}{1+\beta^{\frac{-1}{\gamma}}  \left(1+r_{t+1} \right)^{\frac{\gamma - 1}{\gamma}}}  \\
=& \left(1+a_t \right) \left(1+n_t \right) \left( \frac{\delta + r_{t+1}}{1+r_{t+1}} \right) k_{t+1}  \label{example-dynamics}
\end{align}
where, we used $A_{t+1} = (1+a_t)A_t$, similarly to (\ref{population}).

As we are aware that $f'(k_{t+1})=x_{t+1} =r_{t+1}+ \delta$, and $f'(k_t) = \alpha k_t^{\alpha - 1}$ {{{{stand}}}} for the Cobb-Douglas case, we may specify the above dynamical property of $k_t$ labeled (\ref{example-dynamics}) as follows:
\begin{align}
\frac{(1 - \alpha) k_t^{\alpha}}{1+\beta^{\frac{-1}{\gamma}} \left( 1+\alpha k_{t+1}^{\alpha - 1} - \delta \right)^{\frac{\gamma - 1}{\gamma}}} = \frac{ \left(1+a_t \right) \left(1+n_t \right) \alpha k_{t+1}^{\alpha} }{1+ \alpha k_{t+1}^{\alpha - 1} -\delta}   \label{maindyn}
\end{align}

\subsection{Fertility Rate in the Steady-State}
It is the property of an OLG model under Cobb--Douglas production, as presented above, that the representative agent's utility will be maximized either by a very large or a very small {{{steady-state}}} population growth rate $n$.
In other words, there is such a fertility rate that \textit{minimizes} the utility.\footnote{
This property was shown by \citet{serendipity2} right after \citet{serendipity} studied the serendipity theorem.
On the other hand, endogenous fertility models, that include fertility in the utility function, have interior maximizing solutions \citep{abio, fanti}.
}
That is to say that the fertility rate is not endogenous but rather has to be determined from the planning perspective.
Here, we will see how the fertility rate $n$ and a representative utility $u$ are related by way of the per-capita capital stock $k$ at the steady state, as an intermediary.

First, we solve the steady state of the main OLG dynamics (\ref{maindyn}) for $n$ in terms of $k$.\footnote{
It may be more convenient to solve the opposite, but could not be done analytically.}
Below we assume that $A$ is constant (i.e., $a=0$) for simplicity.
\begin{align*}
\frac{(1 - \alpha) k^{\alpha}}{1+\beta^{\frac{-1}{\gamma}} \left( 1+\alpha k^{\alpha - 1} - \delta \right)^{\frac{\gamma - 1}{\gamma}}} =\left(1+n \right) \frac{ \alpha k^{\alpha} }{1+ \alpha k^{\alpha - 1} -\delta}   %
\end{align*}
Next, we express the steady state utility $u$ in terms of $k$, via the following identities:
\begin{align*}
u&=\frac{c^{1-\gamma}}{1-\gamma}+\beta\frac{d^{1-\gamma}}{1-\gamma},~~~
c=w-s,~~~
d=(1+r), \\
w&= A(1 -\alpha)k^\alpha,~~~
r=\alpha k^{\alpha -1} - \delta,~~~\\
s&=\frac{w}{1+\beta^{\frac{-1}{\gamma}} \left(1+r\right)^{1-\frac{1}{\gamma}}} 
\end{align*}

\begin{figure}[htbp!]
\begin{centering}
\includegraphics[scale=0.43]{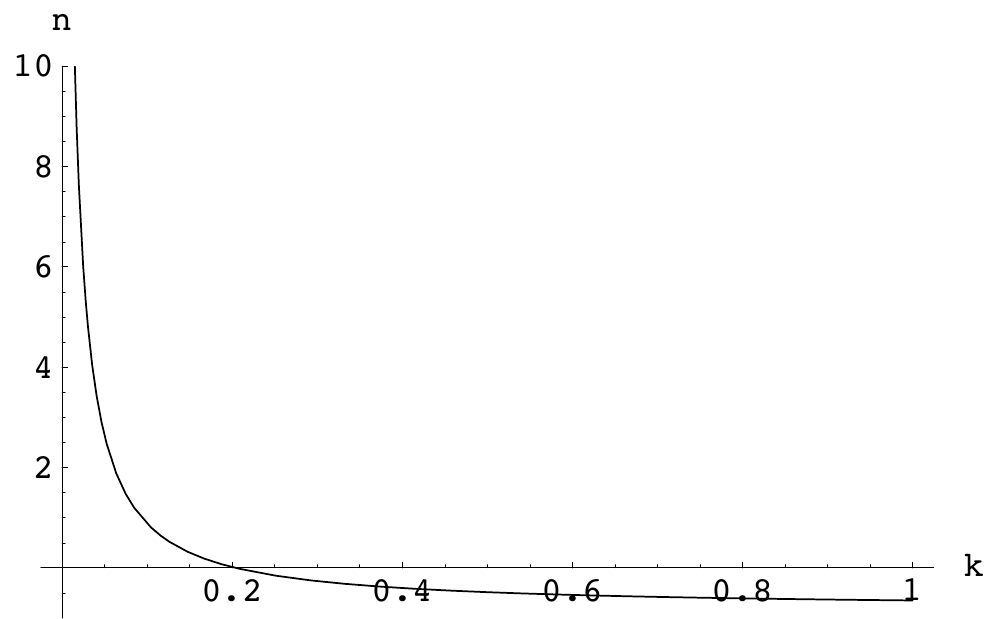}
\includegraphics[scale=0.43]{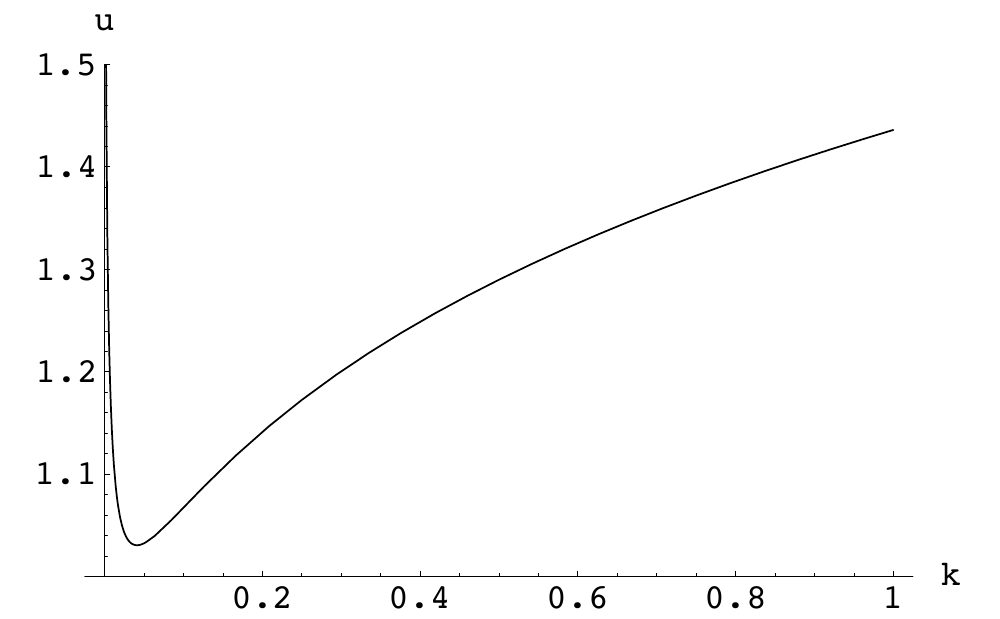}
\caption{Steady state utility level $u$ with respect to the fertility rate $n$ via per-capita capital $k$.}\label{fig_nu}
\end{centering}
\end{figure}

In stead of showing analytically the connection between the fertility rate $n$ and a representative utility $u$, we use a numerical example (with parameters $A=1,\alpha = 0.3, \beta=0.5, \gamma=0.4, \delta=0.3$), which we also will be using in the later demonstration, to show graphically how $n$ affects $k$ and $k$ affects $u$.
As shown in Figure \ref{fig_nu}, $k$ and $n$ are negatively correlated, while a dip exists in the correlation between $k$ and $u$.
So, we find that $n=3.37$ minimizes $u$, while any fertility rate higher or lower than this level will increase the representative utility.
We will use this fertility rate as the upper bound in our later discussions, as fertility must be upper bounded simply from a biological perspective.

\subsection{Energy Efficiency and Resource Availability}
As time proceeds, the cumulation of population adds innovation, thereby augmenting productivity, while at the same time, the resource for energy is consumed.
In this regard, we model the transition of $A_t$ (and thus of $a_t$) with respect to the population schedule $N_t$.
First, we decompose $A_{t}$ into two parts:
\begin{align}
A_{t} = \frac{E_t}{Z_t} \frac{Z_t}{N_t} = G_t H_t,~~~~ A_1=1  \label{twoparts}
\end{align}
where, $E_t = A_t N_t$ denotes the amount of \textit{energy} used in production and $Z_t$ the amount of \textit{resource} exploited in period $t$.
In other words, besides capital, production is carried out by an intermediate factor called energy, which is produced by the primary factor (called labor). 
Further, let us call $G \equiv E/Z$ and $H\equiv Z/N$, the energy efficiency (EE) and the resource availability (RA), respectively.

Assume RA is reduced by the total (cumulated) amount of the resource exploited up until period $t$, while the exploitation of the resource is affected by the RA itself.
That is,
RA is affected negatively by the cumulated $Z_{t} = H_t N_t$ such that:
\begin{align*}
H_{t+1} 
= 1 - \rho \sum_{i=1}^t H_i N_i, ~~~~H_1=1
\end{align*}
where, $\rho > 0$ designates the scale parameter that converts the exploited resource into RA.
By recursive means we will see that the above formula reduces to:
\begin{align}
H_{t+1} = 
\left( 1 - \rho N_t\right) H_t
\label{ht}
\end{align}
For the sake of simplicity, we also assume that EE is enhanced via the magnitude of the population as well as the level of the underlying EE, in the following manner:
\begin{align*}
G_{t+1} =
1 + \sigma \sum_{i=1}^t G_i N_i, ~~~~ G_1=1 
\end{align*}
where, $\sigma >0$, presumably stochastic, designates the scale parameter for innovation in EE.
By recursive means we may also write this as:
\begin{align}
G_{t+1} =
\left( 1 + \sigma N_t\right) G_t
\label{gt}
\end{align}

Combining (\ref{ht}) and (\ref{gt}), we have the following evolution of labor productivity $A_t$ while taking EE and RA into consideration:
\begin{align}
a_t + 1= \frac{A_{t+1}}{A_t} =\frac{G_{t+1}H_{t+1}}{G_{t}H_{t}} = \left( 1 + \sigma N_t\right) \left( 1 - \rho N_t\right)  \label{at}
\end{align}
And finally, the {{{state update}}} of the stock energy is described as follows, with $\theta$ being the per-capita stock energy extraction rate, which we assume is constant throughout the entire process.
\begin{align}
R_{t+1} = \bar{R} - \sum_{i=1}^{t} \theta H_i N_i = R_{t} - \theta H_t N_{t}, ~~~~ R_1 = \bar{R} \label{dep}
\end{align}
Note that $\bar{R}$ designates the ultimate reserve of stock energy at $t=1$.

\subsection{The Objective Function}
We first specify the final state at the planning horizon $T$.
As the final generation $T$ is the last generation to be able to consume the remaining stock energy, the following condition must hold, in regard to (\ref{dep}): 
\begin{align}
R_{T+1} = R_T - \theta H_T N_T = 0 \label{dd}
\end{align}
Naturally, $N_{t} =0$ for all $t \geq T+1$, according to (\ref{dd}), since any $N_t$ cannot take a negative value.
From another perspective, as we are focusing only on the stock energy economy while ignoring the flow energy economy, we are implicitly assuming that $\varepsilon = 0$ (see Figure \ref{fig_intro}).

Then, the finite horizon utilitarian SWF would be the sum of the unweighted lifetime utility of different generations up until this planning horizon:
\begin{align} \label{bentham_swf}
\text{utilitarian SWF} = \sum_{t=1}^T u_t N_t 
\end{align}
The finite horizon maximin SWF would be the utility of the individual (generation) with minimum utility among the generations up until $T$ i.e.,
\begin{align} \label{rawls_swf}
\text{maximin SWF} = \min_{1 \leq t \leq T} u_t
\end{align}

\section{Computational Experiment}
\subsection{The problem formulation}
The type of the problem specified above can be categorized as a dynamical optimization problem with a free-horizon transversality condition. 
In particular, the problem focuses on how to maximize the objective function (\ref{bentham_swf} or \ref{rawls_swf}) subject to the technical constraints of the model viz., (\ref{population}, \ref{maindyn}, \ref{ht}, \ref{gt}, \ref{at}, \ref{dep}) with the transversality condition (\ref{dd}).
We first consider the utilitarian SWF maximization problem from a dynamic programming (DP) perspective since backward induction can be useful for numerically solving the policy function $N_t$ with smoothing interpolation.
However, we find that such an approach would be too costly as there are too many state variables (five, namely, $k, N, G, H, R$) to be handled with an affordable computing device within a decent amount of time, let alone the implicit nature of the main dynamics (\ref{maindyn}). 

We then consider simplifying the main dynamics by choosing log utility ($\gamma =1$) and complete depreciation ($\delta =1$), and further, by saving the state variables through employing $\rho = \sigma = 0$, so that $G$ and $H$ become constant (thus $a=0$).
These settings seem to be meaningful from a computational perspective.
However, log utility can lead to an irrelevant corner solution ($N_t =0$) for the case of utilitarian SWF because the sum of negative utility values (when consumption levels are less than unity) is negative so that the zero population would be better off.\footnote{Log (negative) utility is useful to the extent of relative comparison. 
Absolute assessment, however, may not be appropriate in this case as it presumes the `better never to have been' philosophy \citep{benatar}.}
Moreover, constant productivity (i.e., $a=0$) is hard to accept because our central concern is with the technological progress provided by the dynamics of population. 

As an alternative mean, we consider the problem from a nonlinear programming (NLP) perspective. 
NLP can be applicable to finite horizon dynamical optimization problems such as the current one, and yet can handle many state variables.    
For further investigation, we rewrite our dynamic NLP problem as follows:
\begin{align*}
&\text{Maximize} \, 
\sum_{t=1}^T u_t N_t ~~~~\text{  or  }~~~ \text{Maximize} \,\, u_{\min} \\
&\text{subject to }  \\
& u_t = \frac{c_t^{1-\gamma}}{1-\gamma} + \beta \frac{d_{t+1}^{1-\gamma}}{1-\gamma}, ~~~ c_t = w_t - s_t, \\
&d_{t+1} = (1+r_{t+1}) s_t, ~~~ w_t=A_t ( 1-\alpha) k_{t}^\alpha, \\
&r_{t+1} = \alpha k_{t+1}^{\alpha -1} - \delta, ~~~
s_t = \frac{w_t}{1+\beta^{\frac{-1}{\gamma}} \left(1+r_{t+1} \right)^{1-\frac{1}{\gamma}}},  \\
&\frac{(1 - \alpha) k_t^{\alpha}}{1+\beta^{\frac{-1}{\gamma}} \left( 1+\alpha k_{t+1}^{\alpha - 1} - \delta \right)^{\frac{\gamma - 1}{\gamma}}} = \frac{ \left(1+a_t \right) \left(1+n_t \right) \alpha k_{t+1}^{\alpha} }{1+ \alpha k_{t+1}^{\alpha - 1} -\delta}, \\
&1+ n_t = N_{t+1}/N_t, ~~~ 1+a_t = (1 + \sigma N_t) (1 - \rho N_t), \\
&G_{t+1} = (1 + \sigma N_t) G_t, ~~~H_{t+1} = (1 - \rho N_t) H_t, \\
&A_t = G_t H_t, ~~~ R_{t+1} = R_t - \theta H_t N_t, ~~~R_T - \theta H_T N_T = 0, ~~~\\
&\mu \leq u_{\min} \leq u_t, ~~~ \lambda \leq N_t, ~~~ n_t \leq \omega ~~~\text{(for $t = 1, \cdots, T$)}
\end{align*}
where, $\mu >0$ denotes the minimum utility that a life is worthwhile (i.e., critical level); $\lambda$ denotes the minimum possible population to sustain a generation; and $\omega$ denotes the maximum possible growth rate of population. 
Note that the initial conditions for the state variables, namely, $k_1$, $N_1$, $G_1$, $H_1$, and $R_1$ will be given in advance.
Further, because $N_{T+1} =0$, it must be true that $n_{T}=-1$, $k_{T+1}=\infty$, $r_{T+1}=-\delta$ and thus, $d_{T+1}=(1-\delta)s_T$.
We try to solve this problem for the population schedule $\left( N_1, \dots, N_T \right)$, under the parameters written in the lower case Greek letters that are also given in advance.

\subsection{Problem Solving Artifices}
The grand optimum for $T$, which we denote by $T^*$, can be searched by way of sequential (step-wise) computation.
That is, we execute the NLP optimization over many different horizons namely $T=10, 11, 12, \cdots, 100, 101, \cdots$, and so on, until the grand maximum of the objective SWF is obtained.
The planning horizon will be upper-bounded if $\lambda$, the minimum possible population for a generation, is strictly positive, as each generation must be consuming a certain portion of a limited stock energy.
Hence, by setting a sufficiently large $\lambda$ relative to the stock energy bound $\bar{R}=R_1$, we would be able to have a $T^*$ that is computationally handleable.
For a trial calculation we employed the parameters displayed in Table \ref{tab_param}, where we set $\lambda = 0.1$ and $\mu = 1$.
Moreover, we set initial stock energy to start with $R_1=50$ and the initial per-capita capital to start with $k_1=0.20$.
In addition, we set the upper and lower bound {{{for}}} $k_t$, and lower bounds {{{for}}} $G_t$ and $H_t$, so as to narrow the range of the feasible solution.
A sample executable GAMS/CONOPT code for the above dynamic NLP with $T=40$ is presented in \hyperlink{ap2}{Appendix 2}.

\subsection{Results and Discussions}
We run the two cases whose parameters are summarized in Table \ref{tab_param}.
The differences between the two cases are seen in the parameters that are concerned with productivity and energy extraction, namely, $\sigma$ and $\rho$.
More specifically, scenario (a) has a slower innovation velocity and quicker break in the availability of energy, as regards to (\ref{at}), whereas scenario (b) has the opposite property.
\begin{table}[hbtp]
  \caption{Technical parameters for scenarios (a) and (b).}
  \label{tab_param}
  \centering
\begin{tabularx}{80mm}{cYY} \hline
Scenario	&	(a)	&	(b) \\ \hline
$\alpha$	&	0.3	&	0.3 \\
$\beta$	&	0.5	&	0.5 \\
$\gamma$	&	0.4	&	0.4 \\
$\delta$	&	0.3	&	0.3 \\
$\sigma$	&	0.001	&	0.002 \\
$\rho$	&	0.006	&	0.001 \\
$\mu$	&	1	&	1 \\
$\lambda$ 	&	0.1	&	0.1 \\
$\omega$ 	&	3.37	&	3.37 \\
\hline
\end{tabularx}
\end{table}

\subsubsection{Scenario (a)}
\begin{figure}[htbp!]
\begin{centering}
\includegraphics[scale=0.65]{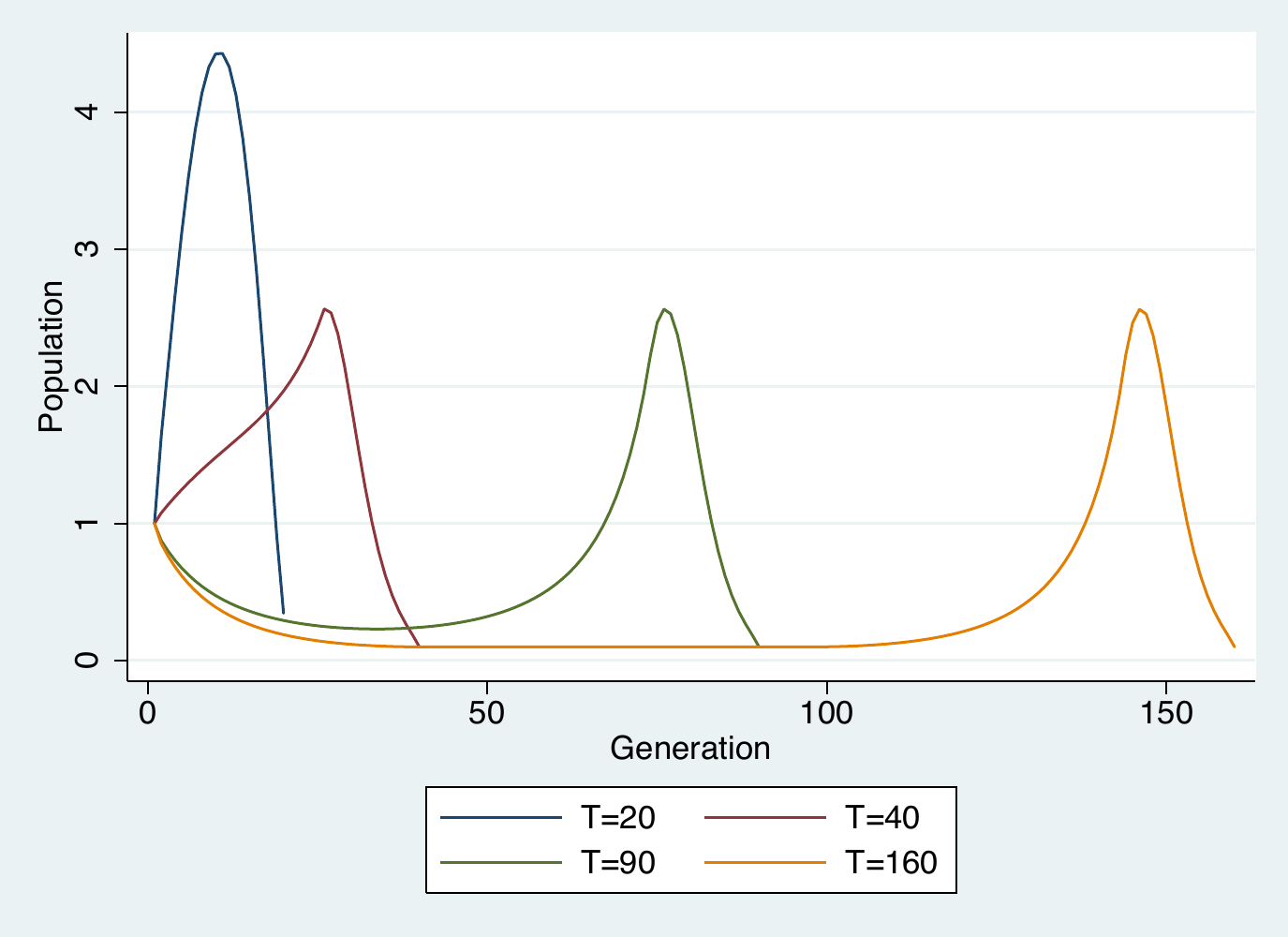}
\caption{Utilitarian optimal population schedule in various planning horizons (scenario (a)).}\label{fig_B_Schedule_a}
\end{centering}
\end{figure}

\begin{figure}[htbp!]
\begin{centering}
\includegraphics[scale=0.65]{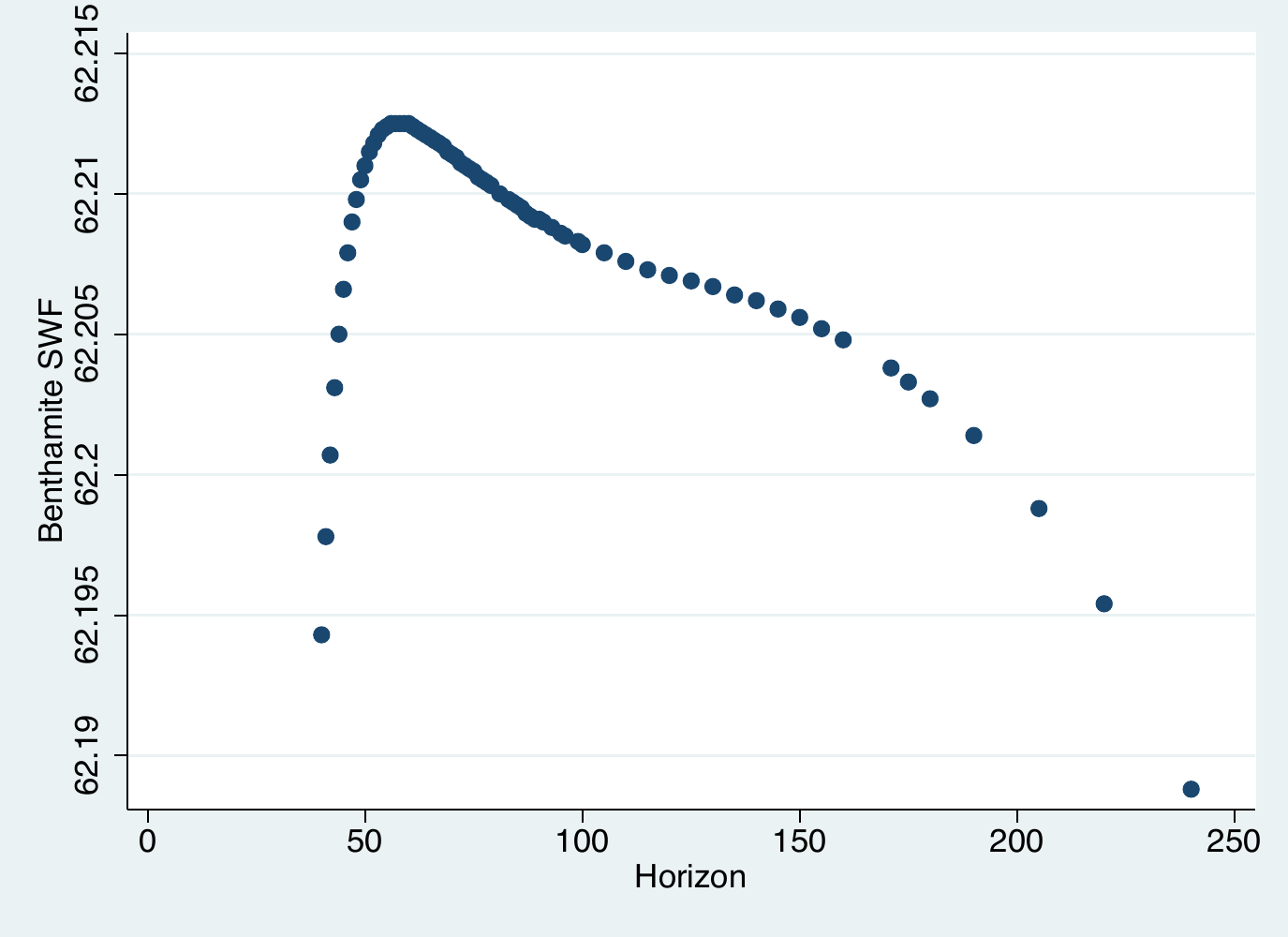}
\caption{Optimal planning horizon search (utilitarian, scenario (a)).}\label{fig_B_horizon_a}
\end{centering}
\end{figure}

\begin{figure}[htbp!]
\begin{centering}
\includegraphics[scale=0.65]{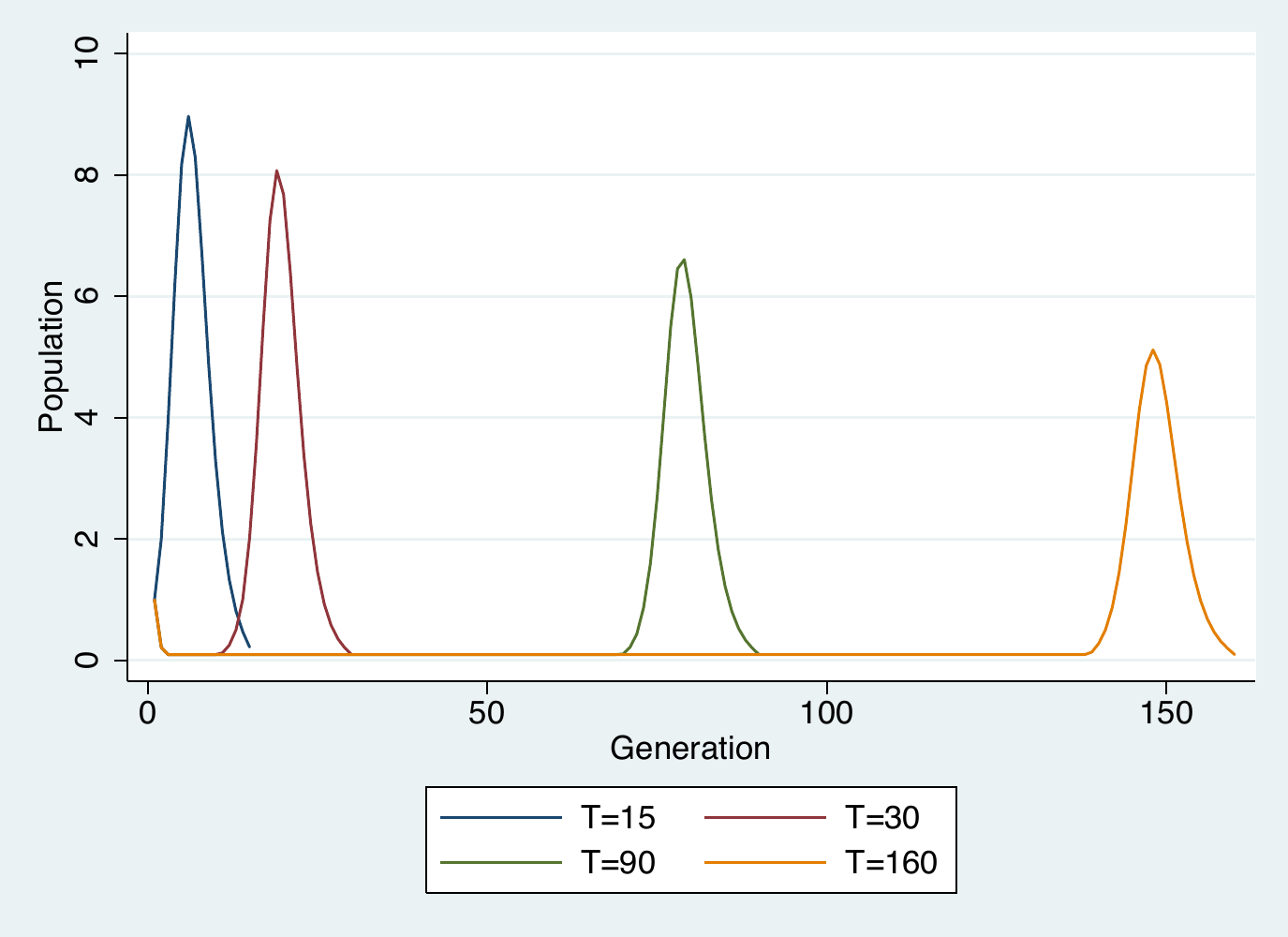}
\caption{Maximin optimal population schedule in various planning horizons (scenario (a)).}\label{fig_R_Schedule_a}
\end{centering}
\end{figure}

\begin{figure}[htbp!]
\begin{centering}
\includegraphics[scale=0.65]{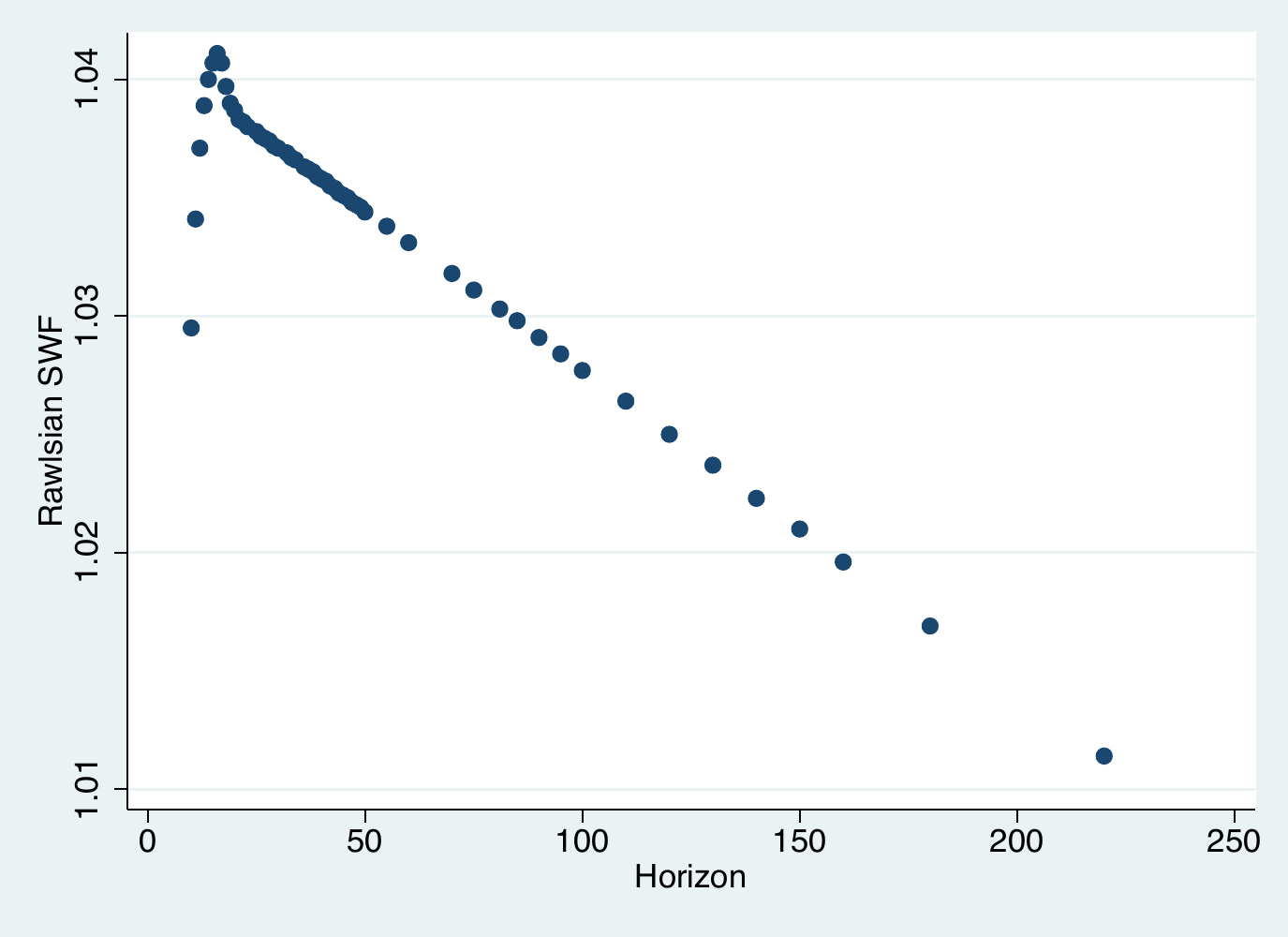}
\caption{Optimal planning horizon search (maximin, scenario (a)).}\label{fig_R_horizon_a}
\end{centering}
\end{figure}

In Figure \ref{fig_B_Schedule_a}, we show the population schedule that maximize utilitarian SWF for the different planning horizons, namely, $T=20$, 40, 90 and 160, under the parameters of scenario (a).
We observe a peak with the same height around the last about 40 generations of the planning horizon.
The grand optimal planning horizon with a utilitarian SWF is at $T^* = 58$, according to Figure \ref{fig_B_horizon_a}.
In Figure \ref{fig_R_Schedule_a}, we show the population schedule that maximize maximin SWF for different planning horizons, namely, $T=15$, 50, 90 and 160.
We observe a peak with descending height around the last 20 or so generations of the planning horizon.
The grand optimal planning horizon with a maximin SWF is at $T^* = 16$, according to Figure \ref{fig_R_horizon_a}.

\begin{figure}[htbp!]
\begin{centering}
\includegraphics[scale=0.65]{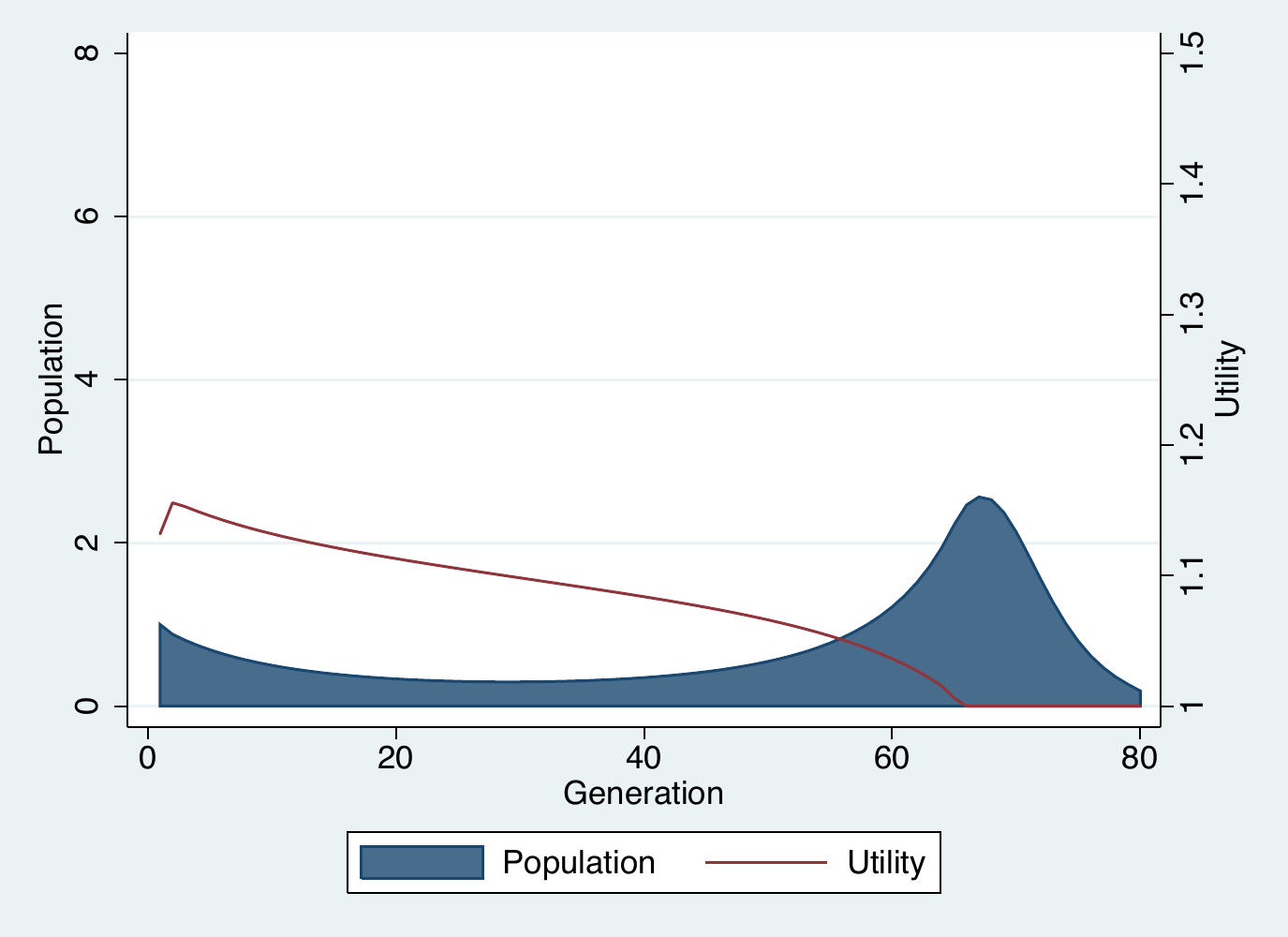}
\caption{Utilitarian optimal population schedule and utility under planning horizon $T=80$ (scenario (a)).}\label{fig_B_Nu80_a}
\end{centering}
\end{figure}

\begin{figure}[htbp!]
\begin{centering}
\includegraphics[scale=0.65]{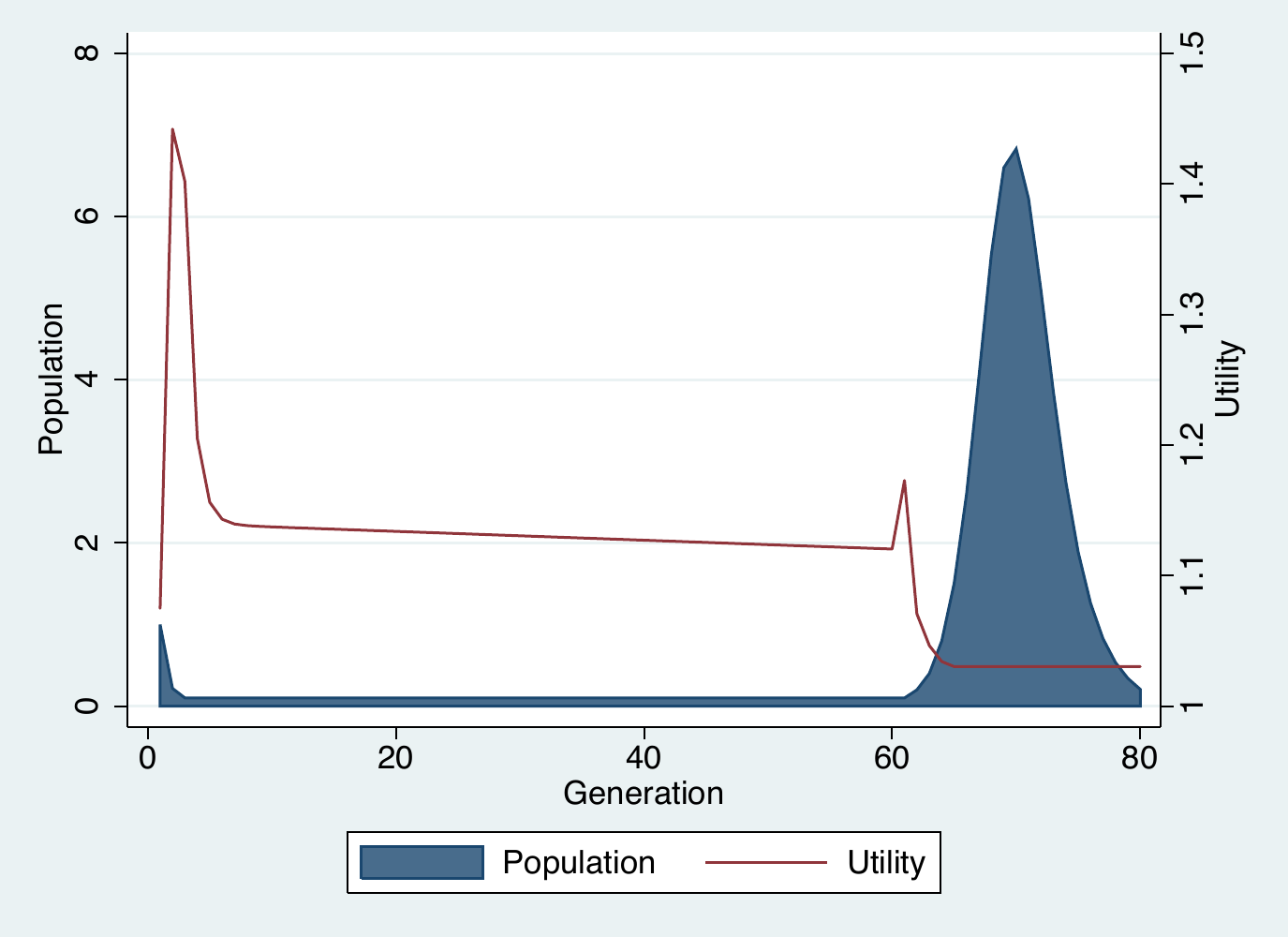}
\caption{Maximin optimal population schedule and utility under planning horizon $T=80$ (scenario (a)).}\label{fig_R_Nu80_a}
\end{centering}
\end{figure}

Figure \ref{fig_B_Nu80_a} and \ref{fig_R_Nu80_a} shows the optimal population schedule, $N_t$, and the corresponding utility schedule, $u_t$, of utilitarian and maximin SWF, respectively, under the planning horizon $T=80$.
Two figures have the same scale measurement, hence they are comparable.
Note that the cumulated total population $\sum_{t=1}^{80} N_t$ was 59.23 for the utilitarian optimum and 58.70 for the maximin optimum.
The maximin SWF i.e., $u_{\min}$ of the utilitarian optimum, when the utilitarian SWF i.e., $\sum_{t=1}^{80} u_t N_t$ was 62.21, was 1.
On the other hand, the utilitarian SWF of the maximin optimum, when the maximin SWF was 1.03, was 61.26.
These tradeoffs between the two {{{assessment criteria}}} will be summarized in Figure \ref{fig_frontier_a}.

\begin{figure}[tbp!]
\begin{centering}
\includegraphics[scale=0.65]{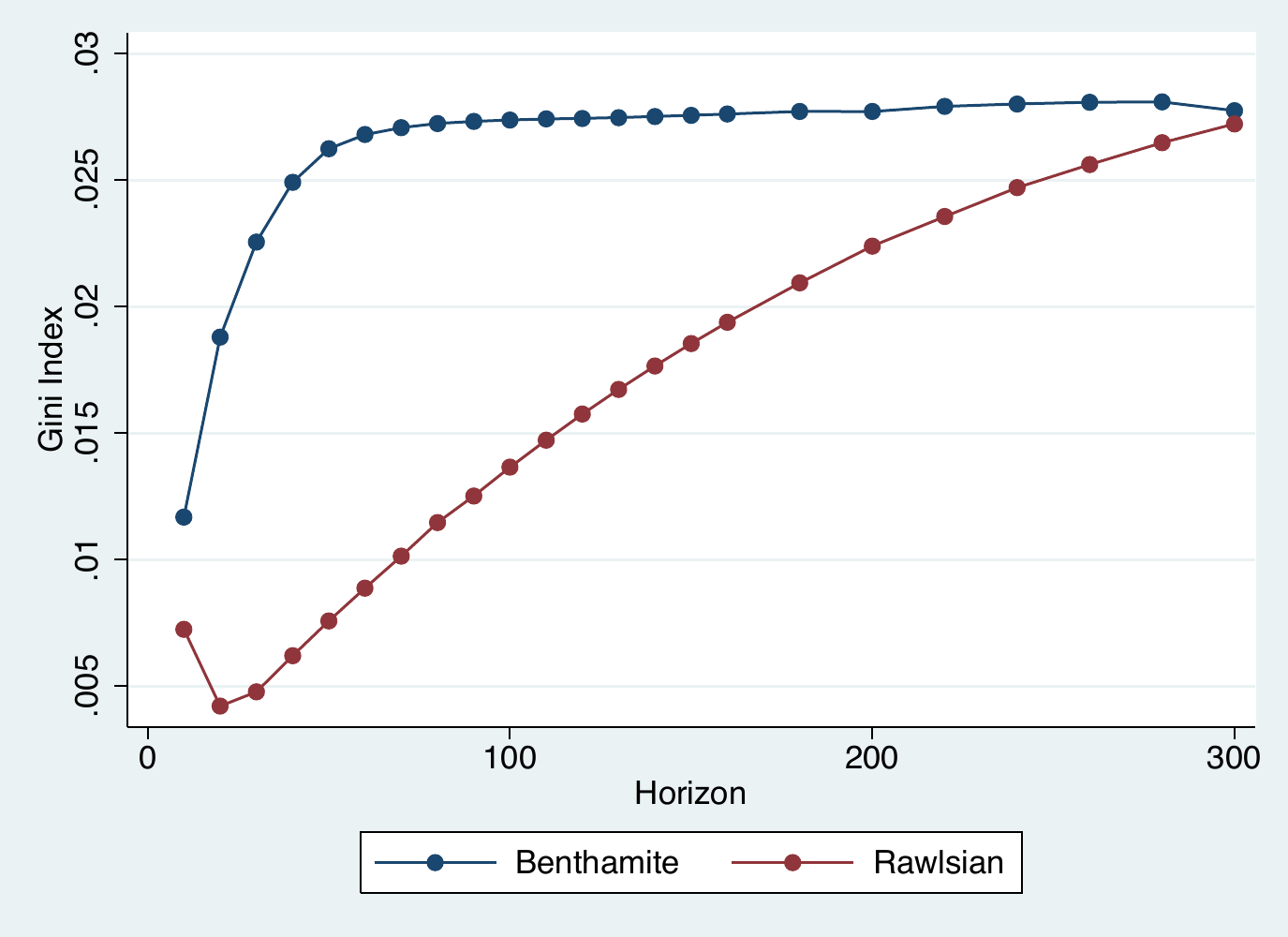}
\caption{Gini indexes (scenario (a)).}\label{fig_gini_a}
\end{centering}
\end{figure}

Also, note that the minimum utility is assigned to the last 15 or so generations of the planning horizon in cases of both the utilitarian and maximin optimum.
Generation-wise utilities seem to be evenly distributed in the case of the utilitarian optimum rather than the maximin optimum, as shown in Figures \ref{fig_B_Nu80_a} and \ref{fig_R_Nu80_a}.
However, the population weighted intergenerational equality (in terms of consequential utility), as measured via the Gini index\footnote{The Gini index is a commonly used measure of inequality, defined by the ratio of the area that lies between the line of equality and the Lorenz curve over the total area under the line of equality \citep[see, e.g.,][]{gini}.
The Gini index can be practically obtained by way of the following formula, using the optimal figures of $N_t$ and $u_t$, for $t=1, \cdots, T$:
\begin{align*}
\text{Gini Index} = 1 - \frac{\sum_{j=1}^T \nu_j 
\left( S_{j-1}+S_j \right)}{S_T}
\end{align*}
where, $S_j = \sum_{i=1}^j \nu_i 
u_i$ and $S_0 =0$; $j$ is the index for generation in utility-increasing order i.e., $u_j < u_{j+1}$; and 
$\nu_j = N_j/\sum_{t=1}^T N_t$.
}, 
indicates the opposite;
the Gini index for the utilitarian optimum population schedule was 0.027 whereas that for the maximin optimum was 0.011, so in this case the maximin optimum is more equitable.
The Gini index for the different planning horizons with scenario (a) is summarized in Figure \ref{fig_gini_a}.

\subsubsection{Scenario (b)}
\begin{figure}[t!]
\begin{centering}
\includegraphics[scale=0.65]{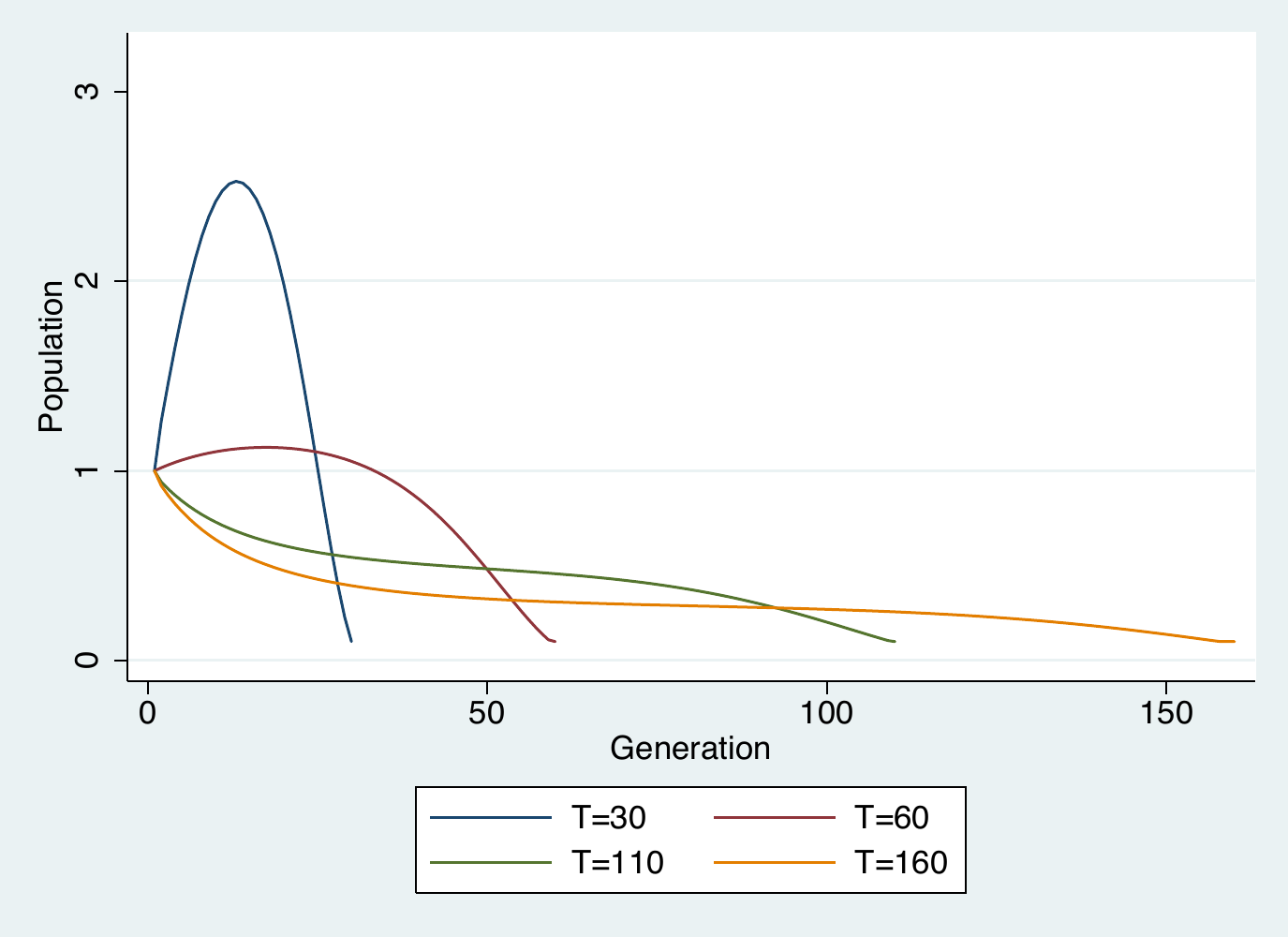}
\caption{Utilitarian optimal population schedule in various planning horizons (scenario (b)).}\label{fig_B_Schedule_b}
\end{centering}
\end{figure}

\begin{figure}[t!]
\begin{centering}
\includegraphics[scale=0.65]{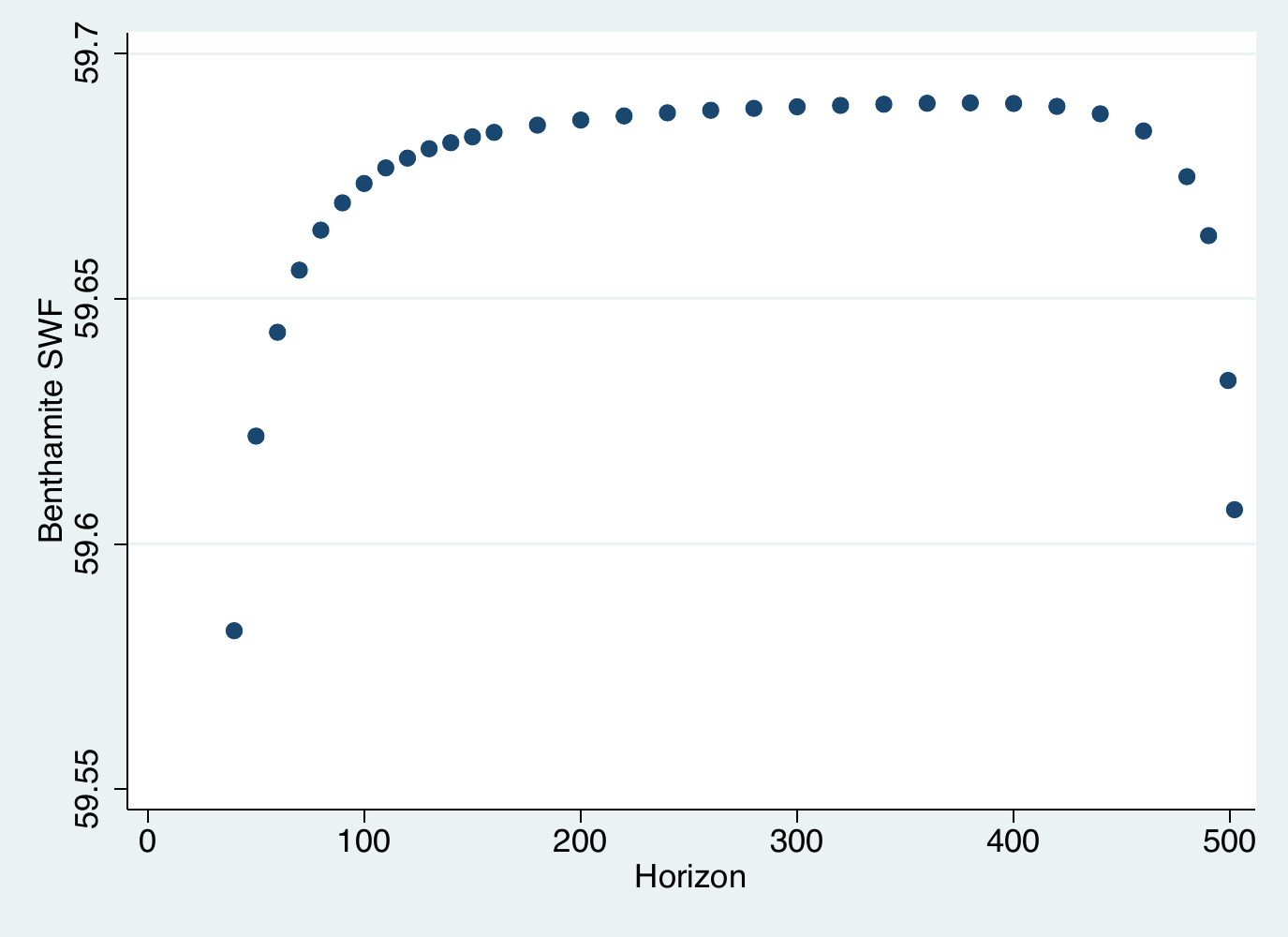}
\caption{Optimal planning horizon search (utilitarian, scenario (b)).}\label{fig_B_horizon_b}
\end{centering}
\end{figure}

\begin{figure}[t!]
\begin{centering}
\includegraphics[scale=0.65]{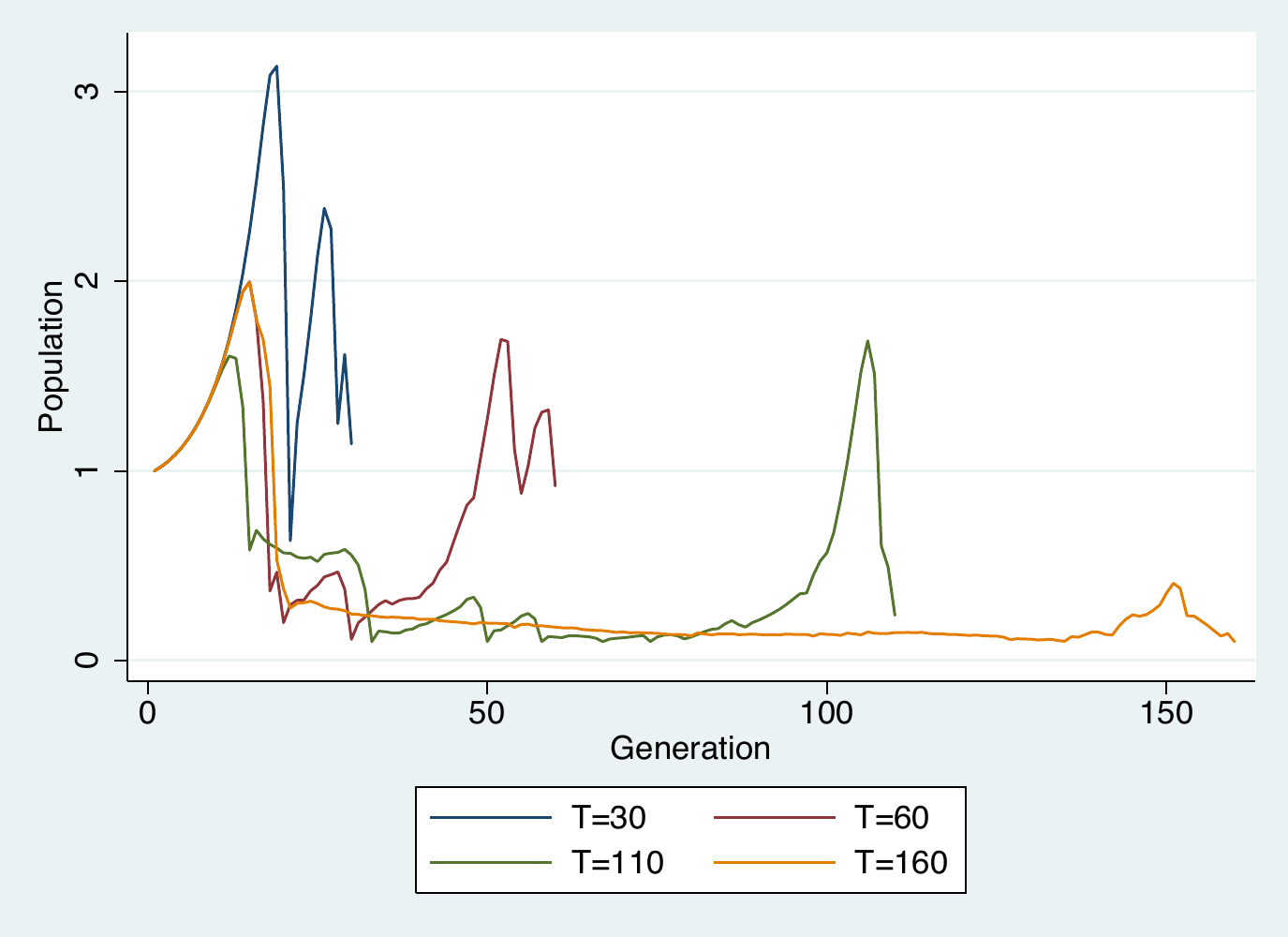}
\caption{Maximin optimal population schedule in various planning horizons (scenario (b)).}\label{fig_R_Schedule_b}
\end{centering}
\end{figure}

\begin{figure}[t!]
\begin{centering}
\includegraphics[scale=0.65]{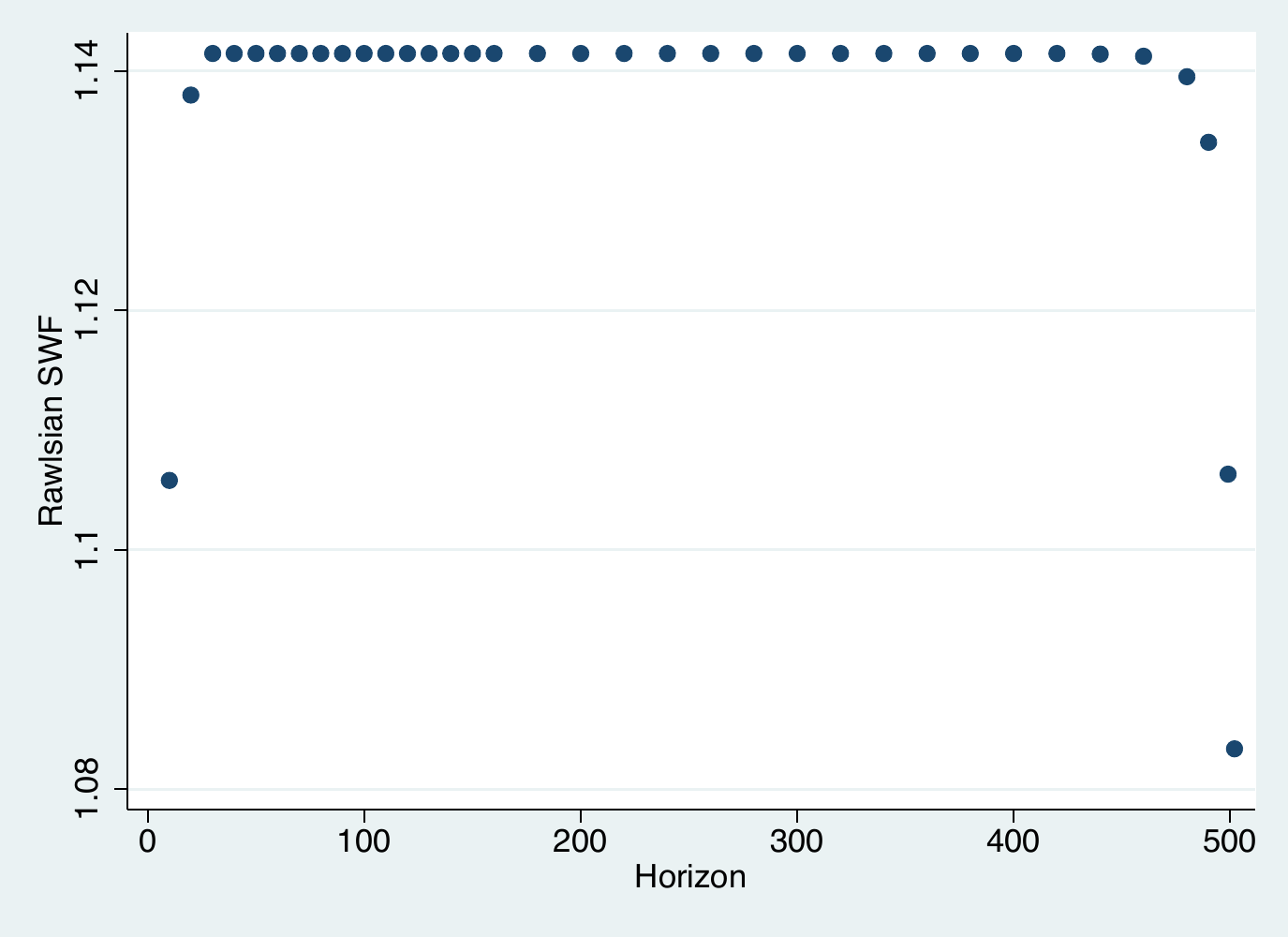}
\caption{Optimal planning horizon search (maximin, scenario (b)).}\label{fig_R_horizon_b}
\end{centering}
\end{figure}

In Figure \ref{fig_B_Schedule_b}, we show the population schedule that maximize utilitarian SWF for different planning horizons, namely, $T=30$, 60, 110 and 160, under the parameters of scenario (b).
A peak will exist until $T=65$, but that will turn into a slope with an inflection point.
The grand optimal planning horizon with a utilitarian SWF is at $T^* = 390$, according to Figure \ref{fig_B_horizon_b}.
In Figure \ref{fig_R_Schedule_b}, we show the population schedule that maximizes the maximin SWF for different planning horizons, namely, $T=30$, 60, 110 and 160.
The result seems to show no definitive tendency, but two peaks may be observable, one in the beginning and another near the horizon.
The grand optimal planning horizon with a maximin SWF is at $T^* = 30$, according to Figure \ref{fig_R_horizon_b}.

\begin{figure}[t!]
\begin{centering}
\includegraphics[scale=0.65]{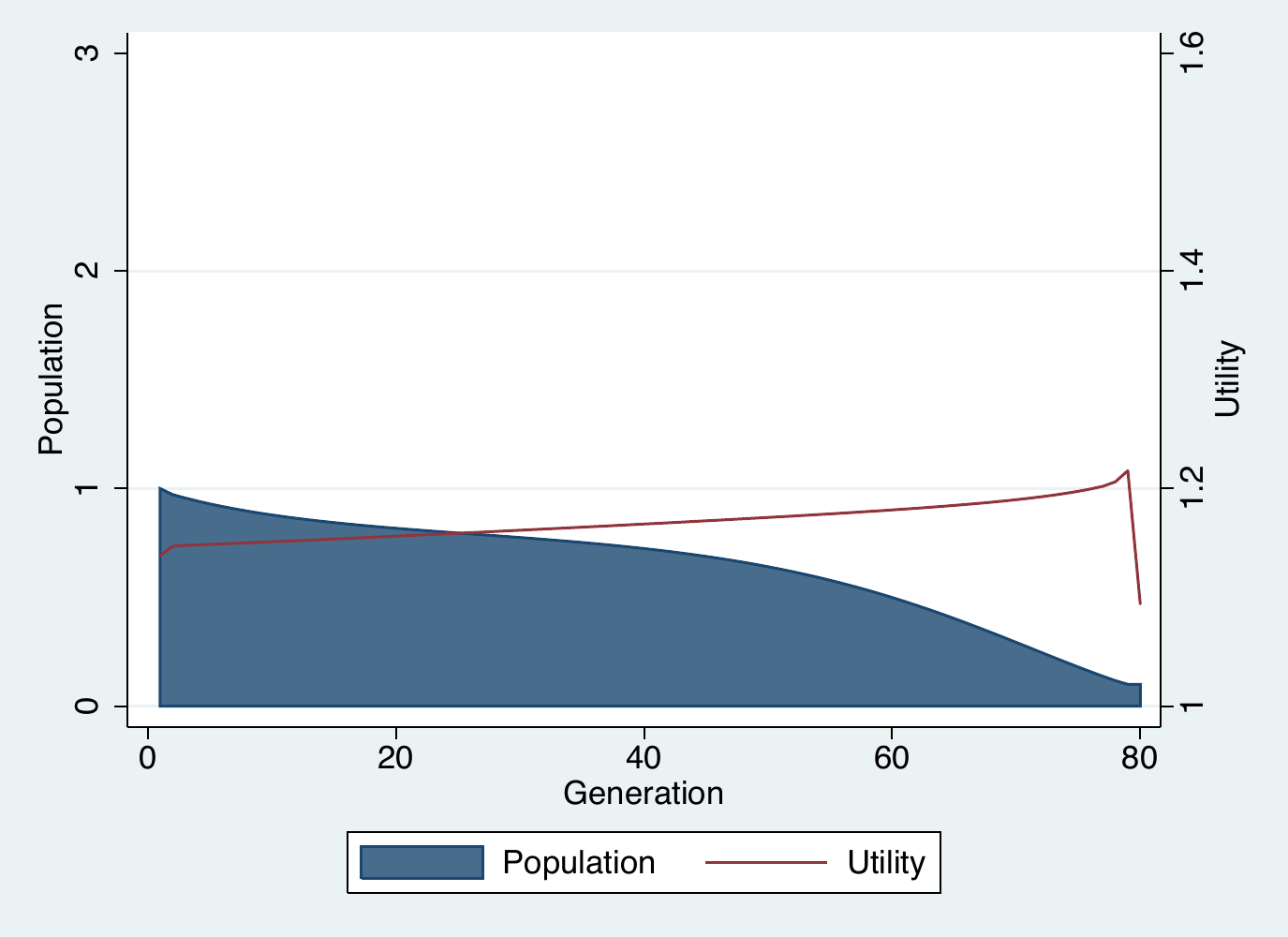}
\caption{Utilitarian optimal population schedule and utility under planning horizon $T=80$ (scenario (b)).}\label{fig_Nu80B_b}
\end{centering}
\end{figure}

\begin{figure}[t!]
\begin{centering}
\includegraphics[scale=0.65]{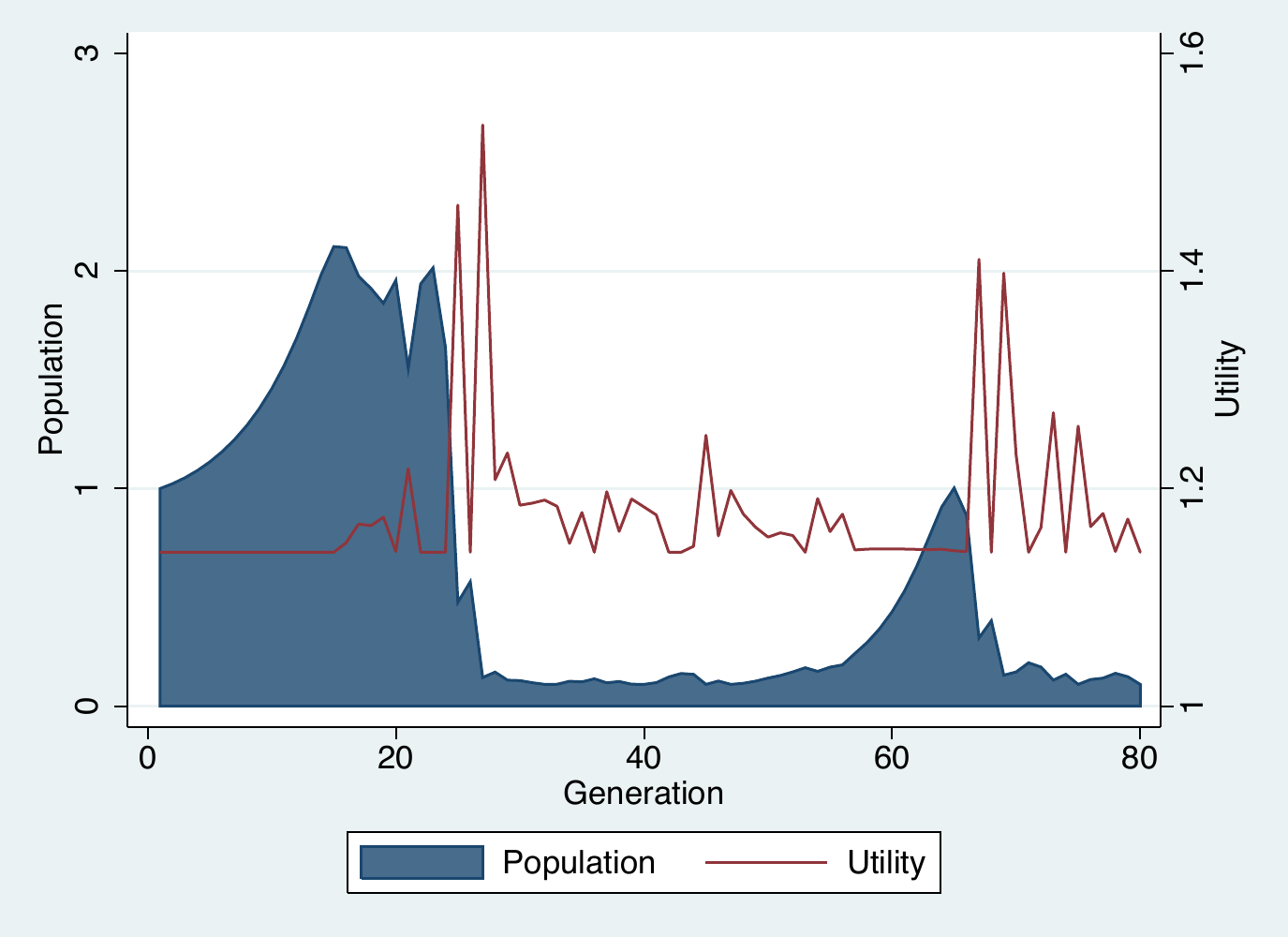}
\caption{Maximin optimal population schedule and utility under planning horizon $T=80$ (scenario (b)).}\label{fig_Nu80R_b}
\end{centering}
\end{figure}

Figure \ref{fig_Nu80B_b} and \ref{fig_Nu80R_b} shows the optimal population schedule, $N_t$, and the corresponding utility schedule, $u_t$, of utilitarian and maximin SWF, respectively, under the planning horizon $T=80$.
Note that the cumulated total population $\sum_{t=1}^{80} N_t$ was 51.27 for the utilitarian optimum and 51.26 for the maximin optimum.
The maximin SWF i.e., $u_{\min}$ of the utilitarian optimum, when the utilitarian SWF i.e., $\sum_{t=1}^{80} u_t N_t$ was 59.66, was 1.137.
On the other hand, utilitarian SWF of the maximin optimum, when the maximin SWF was 1.141, was 59.37.
These tradeoffs will be summarized in Figure \ref{fig_frontier_b}.

\begin{figure}[t!]
\begin{centering}
\includegraphics[scale=0.65]{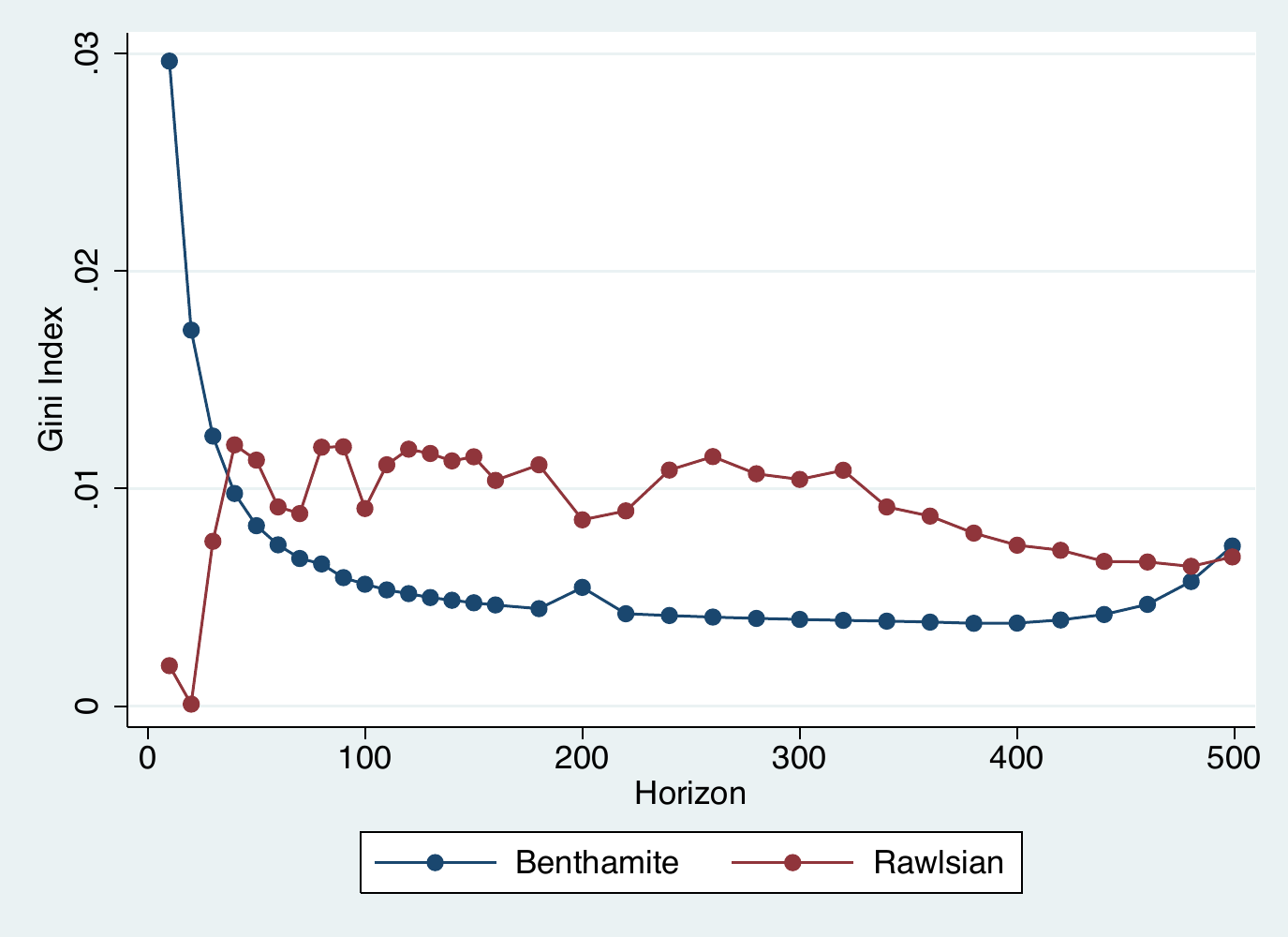}
\caption{Gini indexes (scenario (b)).}\label{fig_gini_b}
\end{centering}
\end{figure}

As we observe in Figures \ref{fig_Nu80B_b} and \ref{fig_Nu80R_b}, utilitarian population schedule and utilities are smoothly distributed within generations whereas they are rather clamorously distributed in the maximin case.
The population weighted generational equality (in terms of utility), as measured via the Gini index for the utilitarian optimum population schedule was 0.0065, whereas that for the maximin optimum it was 0.0120.
Hence in this case utilitarian optimum was more equitable, as can be visually confirmed.
Gini index for different planning horizons with scenario (b) is summarized in Figure \ref{fig_gini_b}.

\subsubsection{Welfare Possibility Frontier}
\begin{figure}[t!]
\begin{centering}
\includegraphics[scale=0.65]{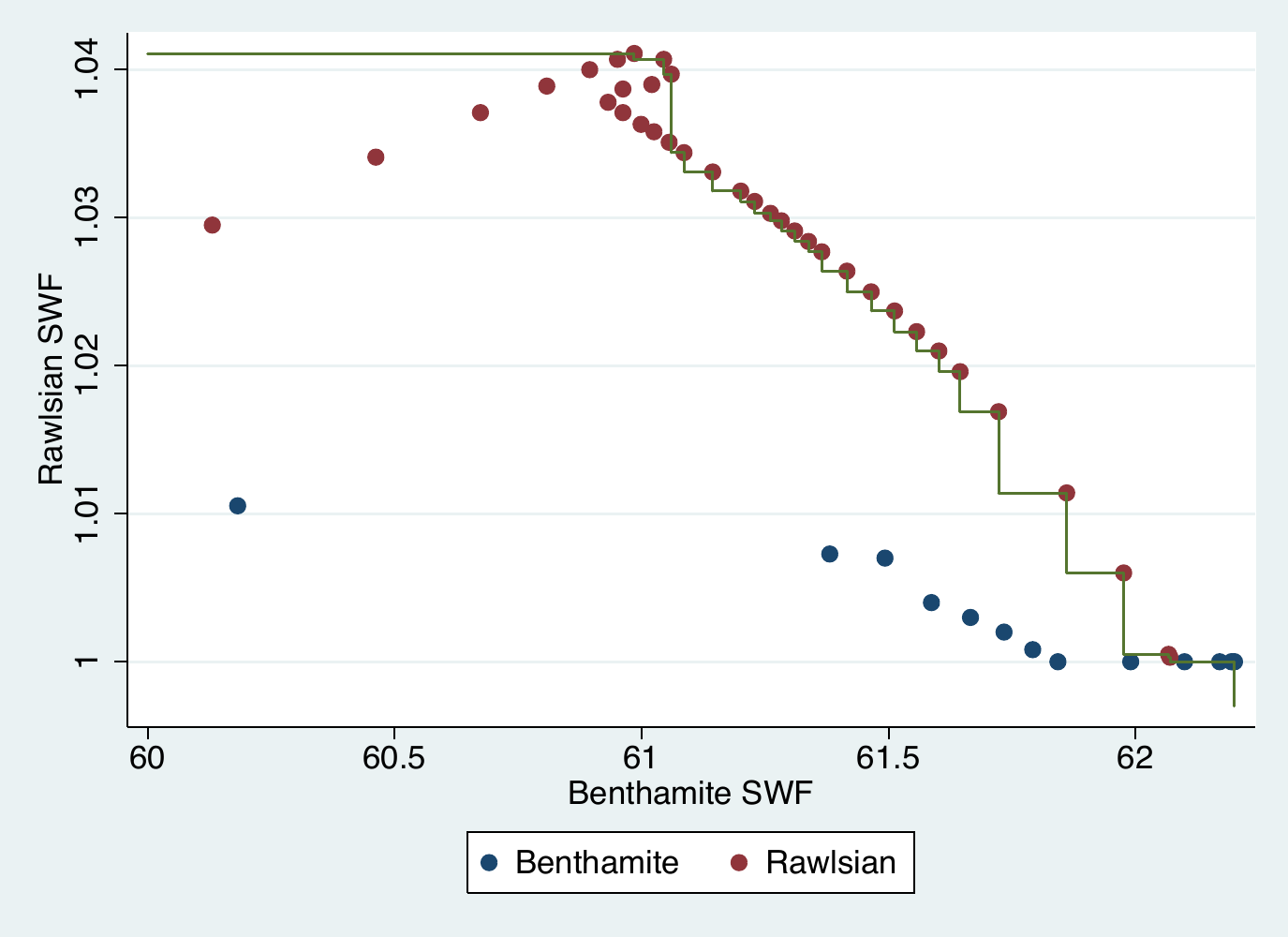}
\caption{Welfare possibility frontier (scenario (a)).}\label{fig_frontier_a}
\end{centering}
\end{figure}

\begin{figure}[t!]
\begin{centering}
\includegraphics[scale=0.65]{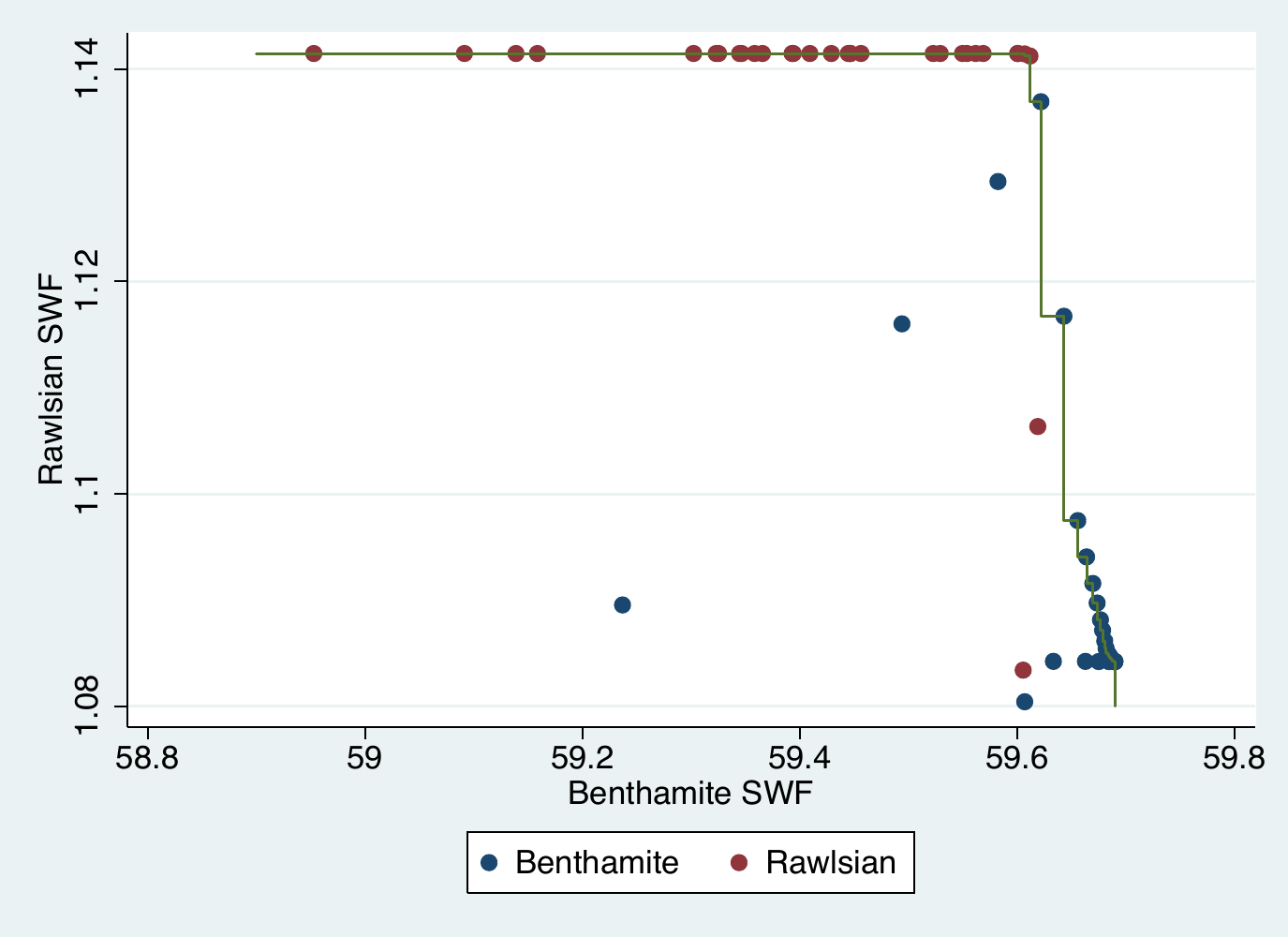}
\caption{Welfare possibility frontier (scenario (b)).}\label{fig_frontier_b}
\end{centering}
\end{figure}

Since any given population schedule, along with the corresponding utility schedule, can be assessed in light of both utilitarian and maximin criteria, we now evaluate the utilitarian optimal schedule in light of maximin criterion, and vice-versa.
Figure \ref{fig_frontier_a} plots the utilitarian maximand on the horizontal axis with the corresponding maximin criterion (i.e., minimum utility) on the vertical axis, in blue dots, whereas the maximin maximand on the vertical axis with the corresponding utilitarian criterion (i.e., total welfare) on the horizontal axis, in magenta dots, all under the parameters for scenario (a).
The same plots under the parameters for scenario (b) are given in Figure \ref{fig_frontier_b}.

The union of the rectangle areas that are hemmed in by every coordinate point (that represents a tuple of possible welfares) in Figures \ref{fig_frontier_a} and \ref{fig_frontier_b} can be considered as the welfare possibility frontier (WPF), representing the potential trade-offs between the two welfare criteria.
Furthermore, observe in Figure \ref{fig_frontier_a} that the utilitarian grad-optimal solution i.e., $(62.213, 1.000)$ at $T^*=58$, and the maximin grand-optimal solution i.e., $(60.985, 1.041)$ at $T^* = 16$ are at the WPF.
Also, in \ref{fig_frontier_b} that the utilitarian grad-optimal solution i.e., $(59.690, 1.084)$ at $T^*=390$, and the maximin grand-optimal solution i.e., $(58.953, 1.142)$ at $T^* = 30$ are at the WPF.

\section{Concluding Remarks}
In this study, we formulated an alternative OLG based dynamic social welfare maximization problem that solves the favorable population schedule of the entire potential human family.
Utilities of the entire potential population, regardless of their generation are valued equally (without discounting) for the case of the utilitarian SWF.
Moreover, because we formulate the problem by way of an NLP framework rather than a backward DP, it becomes quite straightforward to modify the model to solve the case of the maximin SWF.
This approach allowed us to run many NLPs with different  planning horizons, in order to search for the grand optimal planning horizons, under given sets of parameters, for both utilitarian and maximin SWFs.

Population schedules, along with the corresponding utility schedules were obtained for different planning horizons, and intergenerational equities were discussed by way of the Gini indexes obtained from those figures. 
Furthermore, we were able to show the trade-offs between the utilitarian and maximin SWFs, verifying that an SWF maximizing population schedule under a sub-optimal planning horizon can lie at the welfare possibility frontier.
However, we may say nothing definitive about the disposition of the results, as they merely represent a reality given that the parameters were arbitrarily chosen for demonstration purposes.
Even so, our findings that there are no typical pattern for the optimal population schedule, and that they are quite sensitive to the parameters regarding technology, should not be overlooked.

\section*{\hypertarget{ap1}{Appendix 1: Equity and Infinity: An Example}}
Let us consider a macroeconomic dynamical planning problem of the following:
\begin{align}
\max_{c_t \, (t=0,1,\cdots,T)} \sum_{t=0}^T b^t \ln c_t
\text{~~s.t.~~}
k_{t+1} = a k_{t} - c_{t}
\end{align}
where, $c$ denotes consumption (action) and $k$ denotes capital (state), where $k_0$ is given.
The parameter $a$ represents productivity of capital which is assumed to be constant over time, and $b$ represents the discount factor.
The utility (payoff) function is mimicked by a logarithmic function.
Note that population is neglected and that there is only one individual for each generation $t$.

We can solve this problem by backward induction.
First, let us interpret this problem into the following Bellman equation (subscripts are omitted):
\begin{align}
V(k) = \max_c \left\{ \ln c + b V\left( ak - c \right) \right\}
\label{bellman}
\end{align}
Because there will be no $T+1$ generation, it follows that $k_{T+1}=ak_T - c_T= 0$, so that for $t=T$, (\ref{bellman}) becomes
\begin{align*}
&c=ak,
&V(k) = \ln ak
\end{align*}
Next step is to solve (\ref{bellman}) for $t=T-1$, that is to
\begin{align*}
\max_c \left\{
\ln c +b \ln \left(a \left( a k - c \right) \right)
\right\}
\end{align*}
The solution can be found as below:
\begin{align*}
c = \frac{ak}{1+b}
\end{align*}
Accordingly the value function for $t=T-1$ is:
\begin{align*}
V(k) = \ln \left(\frac{ak}{1+b}\right) + b\ln \left(\frac{a^2bk}{1+b}\right)
\end{align*}
Similarly, we solve (\ref{bellman}) for $t=T-2$, that is to:
\begin{align*}
\max_c \left\{
\ln c 
+b \left(\ln \left(\frac{a(ak-c)}{1+b}\right)
+b \ln \left( \frac{a^2 b (ak-c)}{1+b} \right) \right)
\right\}
\end{align*}
The solution can be found as below:
\begin{align*}
c = \frac{ak}{1+b+b^2}
\end{align*}
We may repeat this procedure $T$ times and arrive at the following result:
\begin{align*}
c_0 = \frac{ak_0}{1+b+ \cdots + b^T}
=\frac{1-b}{1-b^{T+1}}ak_0
\end{align*}
Notice that this gives the optimal action (policy) of the current generation $t=0$.
Generation $t$'s optimal consumption $c_t$ can be obtained by recursion, as follows:
\begin{align}
&c_t = \frac{b^{t} (1-b)}{1-b^{T+1}}a^{t+1} k_0, 
&t=0, 1, \cdots, T
\label{solu}
\end{align}

Now, consider pure intergenerational equity that is $b=1$.
In that event (\ref{solu}) becomes indeterminate form, but by virtue of L'H{\^o}spital's rule, we obtain:
\begin{align}
c_t=\lim_{b \to 1} \frac{b^{t} (1-b)}{1-b^{T+1}}a^{t+1} k_0
=\frac{a^{t+1} k_0}{T+1}
\label{lim}
\end{align}
We may then consider what happens if $T\to \infty$.
It is obvious that we have $c_0=c_1=\cdots=0$ as regards (\ref{lim}).
We may interpret this result in words that if we were to maintain equity ($b=1$) in infinity ($T \to \infty$) the entire generations must all be equally non-existent.

\section*{\hypertarget{ap2}{Appendix 2: Sample GAMS Code}}
\lstset{language=GAMS}
\begin{lstlisting}
* OPFH by K.N. 
* Dimension of generation must coincide with the planning horizon tt
* Requires CONOPT3 for a longer planning horizon 
SET
j      generation  / 1*40 /
POSITIVE VARIABLES
G      energy efficiency
H      resource availability
A      labor productivity
N      population 
R      remaining (stock) energy resource
k      per-capita capital
c      per-capita consumption when young
d      per-capita consumption when old
s      savings by a generation
rr     rate of return;
VARIABLES
aa     labor productivity growth rate
nn     population growth rate
u      utility of a generation
uN     population weighted utility
SuN    sum of population weighted utilities
minu   minimum utility;
PARAMETERS
alpha  output elasticity of capital
beta   time preference of individuals
gamma  CRRA parameter (degree of risk aversion)
delta  capital depreciation rate
rho    scale parameter for energy efficiency 
sigma  scale parameter for resource availability 
theta  per-capita stock energy extraction rate
tt     planning horizon
; sigma=0.001; rho=0.006; alpha=0.3; theta=1; beta=0.5; 
  gamma=0.4; delta=0.3; tt=40
EQUATIONS
eq_G, eq_G_ini   state update for G
eq_H, eq_H_ini   state update for H
eq_A, eq_A_dyn   state update for A
eq_N, eq_N_ini   state update for N
eq_R, eq_R_ini   state update for R
eq_k, eq_k_ini   state update for k (main OLG dynamics)
eq_c             determination of c
eq_s, eq_s_fin   determination of s
eq_rr            determination of rr
eq_d, eq_d_fin   determination of d
eq_u             determination of u
eq_uN            determination of uN
eq_Bentham       utilitarian SWF
eq_Rawls;        maximin SWF
eq_G(j+1)..G(j+1)=e=(1+sigma*N(j))*G(j); eq_G_ini..G('1')=e=1;
eq_H(j+1)..H(j+1)=e=(1-rho*N(j))*H(j); eq_H_ini..H('1')=e=1;
eq_A(j+1)..A(j+1)=e=(1+aa(j))*A(j); eq_A_dyn(j)..A(j)=e=G(j)*H(j);
eq_N(j)..N(j+1)=e=(1+nn(j))*N(j); eq_N_ini..N('1')=e=1;
eq_R(j)..R(j+1)=e=R(j)-theta*H(j)*N(j); eq_R_ini..R('1') =e= 50;
eq_k(j+1)..(1-alpha)*((k(j))**alpha)=e=(1+(beta**(-1/gamma))*
    (1-delta+alpha*(k(j+1))**(alpha-1))**((gamma-1)/gamma))*
    (1+nn(j))*(1+aa(j))*(alpha*(k(j+1))**(alpha)/(1-delta+
    (alpha*(k(j+1))**(alpha-1)))); eq_k_ini..k('1') =e= 0.20;
eq_c(j)..c(j)=e=A(j)*(1-alpha)*(k(j))**alpha-s(j);
eq_s(j)$(ord(j)<tt)..s(j)=e=(A(j)*(1-alpha)*(k(j))**alpha)/
    (1+(beta**(-1/gamma))*(1+rr(j+1))**(1-1/gamma));
eq_s_fin(j)$(ord(j)=tt)..s(j)=e=(A(j)*(1-alpha)*(k(j))**alpha)/
    (1+(beta**(-1/gamma))*(1-delta)**(1-1/gamma));
eq_rr(j+1)..rr(j+1)=e=alpha*k(j+1)**(alpha-1)-delta;
eq_d(j)$(ord(j)<tt)..d(j)=e=(1+rr(j+1))*s(j);
eq_d_fin(j)$(ord(j)=tt)..d(j)=e=(1-delta)*s(j);
eq_u(j)..u(j)=e=(c(j)**(1-gamma))/(1-gamma)
          +beta*(d(j)**(1-gamma))/(1-gamma);
eq_uN(j)..uN(j)=e=u(j)*N(j) ;
eq_Bentham..SuN=e=SUM(j, uN(j));
eq_Rawls(j)..minu=l=u(j);
<@\textcolor{blue}{\bfseries MODEL}@> IGEFH / ALL / ;
<@\textcolor{blue}{\bfseries option}@> NLP=CONOPT;
N.lo(j)=0.1; k.up(j)=10000000; k.lo(j)=0.0000001;
nn.lo(j)=-1; nn.up(j)=3.37; G.lo(j)=0.0000001; H.lo(j)=0.0000001;
u.lo(j)=1;
<@\textcolor{blue}{\bfseries SOLVE}@> OPFH USING NLP MAXIMIZE SuN;
<@\textcolor[rgb]{0.5,0.5,0.5}{\itshape *  SOLVE OPFH USING NLP MAXIMIZE minu;
}@>\end{lstlisting}

{\small
\raggedright
\bibliography{mybibfile}
}

\end{document}